\documentclass[aps,prd,twocolumn,nofootinbib,superscriptaddress]{revtex4-2}
\usepackage{graphicx}  
\usepackage{dcolumn}   
\usepackage{bm}        
\usepackage{amsmath,amssymb,amsxtra,bm,epsfig}   
\usepackage{float}
\usepackage[dvipsnames]{xcolor}
\usepackage{xspace}
\usepackage{cancel,soul,ulem}

\usepackage{tikz-feynman}
\usepackage{stmaryrd}
\tikzfeynmanset{compat=1.1.0}
\usetikzlibrary{arrows}

\tikzset{	
	vertex/.style={circle,draw, minimum size=1.5em},	
	edge/.style={->,> = latex'}	
}
\usepackage[bookmarks, breaklinks, colorlinks,urlcolor=blue, citecolor=red, 
linkcolor=blue]{hyperref}
\hypersetup{
	colorlinks=true,
	linkcolor=red,
	filecolor=magenta,
	urlcolor=ForestGreen,
	citecolor=RoyalBlue}

\usepackage[]{caption}

\newcommand{\be}{\begin{eqnarray*}}
	\newcommand{\ee}{\end{eqnarray*}}

\newcommand{\bee}{\begin{eqnarray}}
	\newcommand{\eee}{\end{eqnarray}}
\newcommand{\beeq}{\begin{equation}}
	\newcommand{\eeq}{\end{equation}}

\usepackage{xspace}

\newcommand{\ba}{\begin{array}}
	\newcommand{\ea}{\end{array}}
\newcommand{\bd}{\begin{displaymath}}
	\newcommand{\ed}{\end{displaymath}}
\newcommand{\besub}{\begin{subequations}}
	\newcommand{\eesub}{\end{subequations}}
\newcommand{\bea}{\begin{eqnarray}}
	\newcommand{\eea}{\end{eqnarray}}

\def\q2 {q^2}

\usepackage{tikz}
\usetikzlibrary{arrows,shapes}
\usetikzlibrary{trees}
\usetikzlibrary{matrix,arrows} 				
\usetikzlibrary{positioning}				
\usetikzlibrary{calc,through}				
\usetikzlibrary{decorations.pathreplacing}  
\usepackage{pgffor}							

\usetikzlibrary{decorations.pathmorphing}	
\usetikzlibrary{decorations.markings}
\tikzset{
	vector/.style={decorate, decoration={snake}, draw},
	provector/.style={decorate, decoration={snake,amplitude=2.5pt}, draw},
	antivector/.style={decorate, decoration={snake,amplitude=-2.5pt}, draw},
	fermion/.style={draw=black, postaction={decorate},
		decoration={markings,mark=at position .55 with {\arrow[draw=black]{>}}}},
	fermionbar/.style={draw=black, postaction={decorate},
		decoration={markings,mark=at position .55 with {\arrow[draw=black]{<}}}},
	fermionnoarrow/.style={draw=black},
	gluon/.style={decorate, draw=black,
		decoration={coil,amplitude=4pt, segment length=5pt}},
	scalar/.style={dashed,draw=black, postaction={decorate},
		decoration={markings,mark=at position .55 with {\arrow[draw=black]{>}}}},
	scalarbar/.style={dashed,draw=black, postaction={decorate},
		decoration={markings,mark=at position .55 with {\arrow[draw=black]{<}}}},
	scalarnoarrow/.style={dashed,draw=black},
	electron/.style={draw=black, postaction={decorate},
		decoration={markings,mark=at position .55 with {\arrow[draw=black]{>}}}},
	bigvector/.style={decorate, decoration={snake,amplitude=4pt}, draw},
}

\tikzstyle{block} = [draw, rectangle, 
minimum height=3em, minimum width=6em]

\begin{document}
	\title{Impact of First-order Electroweak Phase Transition on QCD Axion}

	\author{Dipendu Bhandari}
	\email{dbhandari@iitg.ac.in}
	\affiliation{Department of Physics, Indian Institute of Technology Guwahati, Assam-781039, India}
	
	\author{Soumen Kumar Manna}
	\email{skmanna2021@gmail.com}
	\affiliation{Department of Physics, Indian Institute of Technology Guwahati, Assam-781039, India}

	\author{Arunansu Sil}
	\email{asil@iitg.ac.in}
	\affiliation{Department of Physics, Indian Institute of Technology Guwahati, Assam-781039, India}

\begin{abstract} 
The QCD axion addresses the strong CP problem and dark matter via the misalignment mechanism, typically requiring a 
decay constant $f_a\sim \mathcal{O}(10^{12}$ GeV), unless the initial misalignment angle ($\theta_i$) is fine-tuned. 
This work presents a novel approach where the possibility that the QCD axion satisfying the correct relic is extended over a broad range for $f_a\in [10^8, 10^{14}$] GeV without fine-tuning the $\theta_i$, by introducing a new phase of axion oscillation dynamics across the electroweak phase transition (EWPT). 
This mechanism, we call it {\it{recurrent ~misalignment}}, is a result of a non-renormalizable Peccei-Quinn symmetry breaking interaction involving the axion and the sector responsible for making the EWPT of first order. The scenario not only enhances the QCD axion parameter space in terms of its detection possibility, but also provides a unique probe by detectable gravitational waves.		
\end{abstract}
	\maketitle
The QCD axion places itself as a compelling candidate for dark matter (DM), while its origin is intriguingly connected to the resolution of the strong CP problem in QCD as proposed by Peccei and Quinn (PQ) \cite{Peccei:1977hh,Peccei:1977ur,Weinberg:1977ma,Wilczek:1977pj}. The strong CP problem revolves around the existence of a CP violating $\theta$-term in QCD Lagrangian (sizeable from theoretical perspective), which is severely constrained, $\theta < \mathcal{O}(10^{-10})$, by the experimental limits on neutron electric dipole moment \cite{Crewther:1979pi,Abel:2020pzs}. The PQ solution promotes this $\theta$ to a dynamical field, $\theta(x) = a(x)/f_a$ where $a(x)$ is the Nambu-Goldstone boson (the QCD axion) associated to the spontaneous breaking of global $U(1)_{PQ}$ symmetry by a PQ scalar $\Phi = \left(\eta + f_{a}\right) e^{i\theta}/{\sqrt{2}}$ with breaking scale (or decay constant) $f_a$. The symmetry is also explicitly broken by the axial anomaly during the QCD phase transition, generating a potential for the axion field, which not only generates a tiny mass ${m_{a0}}$ for the axion but also resolves the strong CP problem dynamically as the axion field sets in at the CP-conserving minimum $\theta(x) = 0$ of the potential.

While elegantly solving the strong CP problem, such a QCD axion, if light enough, can be very stable and hence, play the role of dark matter. Cosmologically, the axion field is expected to be effectively frozen at a non-zero initial field value $a_i$ after inflation (without any initial velocity in the so called `misalignment' mechanism \cite{Preskill:1982cy,Abbott:1982af,Dine:1982ah,Turner:1983he,Arias:2012az}), until its mass exceeds the Hubble expansion parameter $(\mathcal{H})$ of the Universe. Consequently, the axion field starts to oscillate coherently around the CP-conserving minimum of the QCD-induced potential and its energy density ($\rho_a$) evolves like non-relativistic matter ($\rho_a\propto R^{-3}$, with $R$ being the scale factor of the Universe).  

Since the onset of oscillation of the QCD axion is crucially dependent on $m_{a0}$ or in other words, on $f_a$
($m_{a0}$ carries inverse dependence on $f_a$ for QCD axion), it is important to know the allowed range of $f_a$ phenomenologically. The only stringent constraint on $f_a$ follows from the observation of SN1987A \cite{Raffelt:2006cw} and neutron star cooling \cite{Caputo:2024oqc}, leading to a lower bound on $f_a$ as $f_a \gtrsim 10^8$ GeV (irrespective of whether the QCD axion be DM or not), which in turn puts an upper limit on the axion mass as $m_{a0} \lesssim 60$ meV. There is as such no upper (lower) limit for $f_a$ ($m_{a0}$) exists. 

Now for the QCD axion to be DM, with the initial misalignment angle $\theta_i=a_i/f_a\sim \mathcal{O}(1)$ as a natural choice, the decay constant turns out to be $f_a \sim 10^{12} ~{\rm GeV}$ (which corresponds to the axion mass in the ballpark of $m_{a0}\sim 5~\mu$eV) to ensure the DM relic satisfaction. Above such $f_a$, the relic satisfaction requires fine-tuned initial condition for $\theta_i$ unless some unconventional axion dynamics as in \cite{Co:2018phi}, or a modification of QCD confinement scale \cite{Heurtier:2021rko} is incorporated. The under-abundance of QCD axion as DM for $f_a < 10^{12}$ GeV can be circumvented (keeping $\theta_i\sim \mathcal{O}(1)$) through mechanisms like kinetic misalignment \cite{Co:2019jts,Chang:2019tvx} (considering an initial non-zero $\dot{\theta}$ contrary to `misalignment' mechanism) and parametric resonance \cite{Co:2020dya,Co:2020jtv,Eroncel:2024rpe} or a scenario where the axion first gets stuck in a wrong minimum before oscillating around the correct minimum, known as trapped misalignment \cite{Nakagawa:2020zjr,DiLuzio:2021gos,DiLuzio:2024fyt}. Moreover, there are several other scenarios that modify the QCD axion dark matter parameter space as in refs. \cite{Harigaya:2019qnl,Barman:2021rdr,Arias:2022qjt,Papageorgiou:2022prc,Xu:2022yxr,Choi:2022btl,Eroncel:2022vjg,Allali:2022yvx,Cyncynates:2023esj,Xu:2023lxw,Lee:2024oaz,Barman:2023icn,Banerjee:2024ykz,Eroncel:2025qlk,Banerjee:2025kov,Lyu:2025jge}. 

\begin{figure}[h]
	\includegraphics[width=1\linewidth]{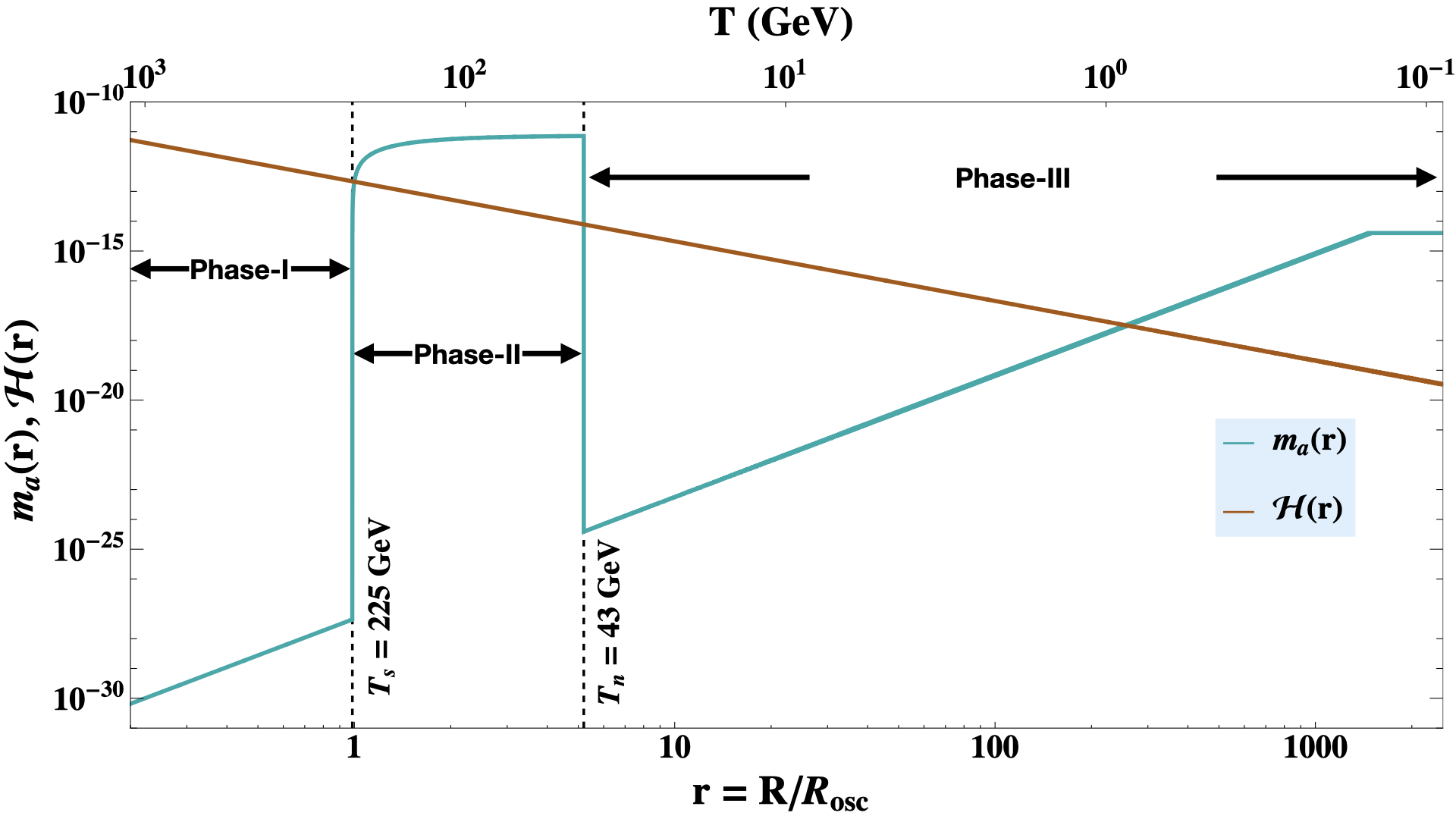}
	\caption{{\small{Variation of axion mass and Hubble around EWPT.}}}
	\label{fig:ma-H}
	\end{figure}
In this work, we propose that, in case the electroweak phase transition (EWPT) is of first order and an explicit PQ symmetry breaking portal interaction between $\Phi$ and the sector responsible for making the EWPT of first order becomes operative, the standard cosmological evolution of the QCD axion gets affected non-trivially during the era of electroweak symmetry breaking
In particular, this results into an early era of axion oscillation during EWPT, an intermittent stoppage of the same (following the completion of EWPT) and a recurrence of the oscillation below the EW scale. However, during the interval between two oscillatory phases, the axion continues to alter its field space, as a consequence of which the earlier misalignment becomes modified prior to the second phase of oscillation. We call it as {\it{recurrent ~misalignment}} mechanism which leads to the possibility of realizing an extended mass (or $f_a$ covering: $10^{8-14}$ GeV) range for the QCD axion as viable DM (instead of $m_{a0} \sim$ 5~$\mu$eV only) without the requirement of any fine tuning for $\theta_i$. Such a broadening of QCD axion mass range as dark matter would be extremely beneficial for detection perspective as many experiments are operating or being designed for future operation in this broader mass range. Alongside, the proposal carries more phenomenological significance as the involvement of FOPT can also generate a potentially detectable stochastic gravitational wave (GW) background. 

Before entering into the core of the proposal, we first note that inclusion of a real SM singlet (odd under a $Z_2$ symmetry) scalar field ($S$) can make the EWPT of first order~\cite{Espinosa:2011ax} through its Higgs ($H$) portal interaction $\lambda_{hs} |H|^2 S^2$. This term generates a tree-level potential barrier after $S$ gets $\it{vev}$ at a temperature, say $T_s$, with appropriate choice of the portal coupling $\lambda_{hs}$, between the electroweak symmetry preserving [$\langle H \rangle = 0; \langle S \rangle = v_s$] and breaking [$\langle H \rangle = v_h; \langle S \rangle = 0$] vacua, enabling the FOEWPT. 
At this moment, it is interesting to observe that the $\it{vev}$ of $S$ can alter the axion's mass in case they are coupled. In order to implement this, we incorporate the following dimension-6 explicit PQ symmetry breaking operator that couples the axion to $S$, given by\footnote{In ref. \cite{Manna:2023zuq}, we previously studied a similar operator, considering the SM Higgs boson $H$ in place of $S$. However, such setup results in a crossover transition and is therefore not associated with a FOPT.}
\begin{align}
	\frac{S^4}{\Lambda^2} \Phi^2   e^{i\alpha} + h.c \,.
	\label{s-portal-int}
\end{align}
The origin of such operator is related to the standard consideration that any global symmetry should anyway be broken by gravity \cite{Giddings:1988cx,Coleman:1988tj,Rey:1989mg,Abbott:1989jw,Akhmedov:1992hh,Kamionkowski:1992mf,Kallosh:1995hi}. Along the line \cite{Draper:2022pvk,Cordova:2022rer}, in the context of weak gravity conjecture, such a breaking may take place even below the Planck scale ($M_{Pl}$), which we consider here by introducing the cut-off scale $\Lambda < M_{Pl}$. Usually, such explicit PQ breaking operator may destabilize the solution of the strong CP problem by PQ mechanism. However, in our scenario, the $v_s$ finally relaxes to origin at the end of FOEWPT, ensuring the QCD vacuum remains CP-conserving one. There can be other higher dimensional operators which are not phenomenologically important for this analysis and hence, corresponding coefficients can be set to zero without loss of generality. We specify them in Appendix~\ref{ap:UV-completion} where a UV completion is also exercised. 

Such a dimension-6 operator of Eq.~{\ref{s-portal-int}} modifies the axion mass $m_{a0}$ (followed from the instanton induced potential $V_0^{\rm QCD}(\theta) =m^2_{a0}(T) f_a^2\left[ 1- \cos\left(\theta\right)\right]$)
during the EWPT as depicted schematically in Fig.~\ref{fig:ma-H}, hence alters  the oscillation dynamics of QCD axion than the standard case, carrying significant impact on the DM relic abundance. To study this, we divide the scenario in three stages: \\
\noindent {\bf{I}}) {\it Above the temperature $T_s$}: during this era, both the Higgs and $S$ fields have vanishing $vev$s, $\langle H \rangle =0, \langle S \rangle =0$ and hence, the axion mass, $m_{a0}(T)$, is solely determined by the non-perturbative QCD effects.\\
\noindent {\bf{II}}) {\it Between $T_s$ and the electroweak nucleation temperature $T_n$}: the $S$ field acquires a non-zero (temperature dependent) $vev$: $\langle S \rangle = v_s(T)$ at $T_s$, leading to an additional contribution $m_{as}^2$ to the axion mass squared: $m_a^2(T)=m_{a0}^2(T) + m_{as}^2$ following Eq.~\ref{s-portal-int}. \\
\noindent {\bf{III}}) {\it Below $T_n$}: the $S$ field's $vev$ now vanishes again, restoring the axion mass to its original value, $m_{a0}(T)$. \\
Hence, the variation of QCD axion's mass across the EWPT can be summarised as 
\begin{widetext}
	\begin{equation}
		m_a (T) =
		\begin{cases}
			m_{a0} \left(\frac{T}{T_{\rm QCD}}\right)^{-b}& \text{;} \quad T>T_s, \\
			\left[ \left(m_{a0}\right)^2 \left(\frac{T}{T_{\rm QCD}}\right)^{-2b} + \frac{4 v_s^4(T)}{\Lambda^2}\right]^{1/2} & \text{;}  \quad  T_n < T \leq T_s,\\
			m_{a0} \left(\frac{T}{T_{\rm QCD}}\right)^{-b} & \text{;}  \quad T_{\rm QCD} < T\leq T_n,\\
			m_{a0} & \text{;}  \quad T\leq T_{\rm QCD},
		\end{cases}
		\label{eq:mass-variation}
	\end{equation}
\end{widetext}
where $m_{a0}=\frac{m_{\pi}f_{\pi}}{f_a} \frac{\sqrt{z}}{1+z}$ is the mass of the axion generated through the non-perturbative QCD effect, with $m_{\pi}$ and $f_{\pi}$ denoting the mass and decay constant of the pion, respectively. Here, $z$ ($\equiv m_u/m_d \simeq 0.48$) represents the ratio of the up to down quark masses, and $b \simeq 4.08$ denotes an index taken from the dilute instanton gas approximation~\cite{Borsanyi:2016ksw}. $T_{\rm QCD}\simeq 150$ MeV~\cite{DiLuzio:2020wdo} is the critical temperature of the QCD phase transition. 
Note that the crucial modification over the standard variation appears through the $4 v_s^4(T)/\Lambda^2$ term in our scenario which plays a pivotal role in the dynamics of  QCD axion. We consider $\alpha = \pi$ in order not to disturb the CP conserving minimum of the axion potential from $\theta = 0$ \cite{Manna:2023zuq}, during  $T_n < T \leq T_s$.

To study the axion dynamics with such $m_a(T)$ variation, it is important to precisely explore the temperature dependence of $v_s$ as well as the nucleation temperature $T_n$ till when the additional axion-mass term $m_{as}$ contributes. For this purpose, we now delve into the details of EWPT employing the tree-level $Z_2$-symmetric potential in the sector responsible for inducing the EWPT of first order, involving $H$ and $S$, as given by~\cite{Espinosa:2011ax}:
\begin{align}
	V_0(H,S) =\, & \mu_h^2(H^{\dagger} H) + \lambda_h(H^{\dagger} H)^2 \nonumber \\ 
	&  + \frac{\mu_s^2}{2} S^2 + \frac{\lambda_s}{4} S^4 + \frac{\lambda_{hs}}{2} (H^{\dagger} H) S^2 , \label{tree-level potential}
\end{align}
where, $\mu_h^2, \mu_s^2 <0$. In addition, thermal correction needs to be included resulting into (detailed in appendix~\ref{ap:effective-potential}) an effective potential $V_{\rm eff} =  V_0 + V_{T}$:
\begin{align}
	V_{\rm eff}(h,S,T) =\, &  \frac{1}{2}( \mu_h^2 + c_h T^2) h^2 + \frac{\lambda_h}{4} h^4 + \nonumber \\
	& \frac{1}{2}( \mu_s^2 + c_s T^2) S^2 + \frac{\lambda_s}{4} S^4 +  \frac{\lambda_{hs}}{4} h^2 S^2,     \label{effective potential}
\end{align}
where, $h$ is the classical background of the Higgs field and 
\begin{align}
	c_h  = \,& \frac{1}{48} (9 g_{\rm w}^2 +3 g_Y^2 +12 y_t^2 + 24 \lambda_h + 2 \lambda_{hs}), \nonumber \\
	c_s =\, & \frac{1}{12} (3 \lambda_s + 2 \lambda_{hs}) .  \label{ch-cs}
\end{align} 
In constructing the effective potential (Eq.~\ref{effective potential}), we adopt high-temperature approximation and retain only the leading-order $T^2$ thermal corrections, neglecting the full one-loop Coleman–Weinberg contributions beyond this.


In our setup, the phase transition proceeds in two steps. At a reasonably high temperature, all symmetries (electroweak as well as $Z_2$ symmetry) are restored, and both the Higgs and the $S$ field have zero $vev$s. Once the Universe cools down to $T_s$, the $S$ field acquires a temperature-dependent $vev$ (breaking the $Z_2$ symmetry), 
\begin{align}
	\langle S \rangle= v_{s}(T) =\sqrt{ \frac{-\left( \mu_s^2 + c_{s} T^2\right)}{\lambda_s}}, \label{s-vev}
\end{align}
while the Higgs $vev$ remains zero at this stage. This transition, from $\langle S \rangle= 0$ to $\langle S \rangle\neq 0$ at $T_s$, can occur via a simple crossover or a second-order phase transition. As the temperature decreases further, a new minimum with $\langle h \rangle = v_h(T) = v_0 \sqrt{\left( 1 - 2 \frac{c_h}{m_h^2} T^2 \right)}\neq 0, \langle S \rangle= 0$ appeares. At a certain critical temperature, $T_c$ ($<T_s$), this minimum becomes degenerate with the former one  $\left( \langle h \rangle = 0, \langle S \rangle \neq 0 \right)$, separated by a potential barrier. Here, $v_0 =246$ GeV and $m_h=125$ GeV denote the zero-temperature SM Higgs $vev$ and Higgs mass respectively. Below $T_c$, the EWSB minimum $\left(\langle h \rangle= v_h(T), \langle S \rangle= 0 \right)$ emerges as the global (or true) vacuum. 

Although the true-vacuum bubbles begin to form at $T_c$, the bubble nucleation, {\it i.e.}, the efficient conversion of symmetric phase ($\langle h \rangle = 0, \langle S \rangle = v_s(T)$) of the Universe to broken one  $\left(\langle h \rangle =v_h(T), \langle S \rangle= 0 \right)$, remains suppressed due to the small false-vacuum decay rate, $\Gamma_d(T)$ than the expansion rate of the Universe $\mathcal{H}(T)$. Below $T_c$ at a certain temperature, when the probability of forming at least one bubble per horizon volume becomes $\mathcal{O}(1)$, the transition from the false to the true vacuum effectively
proceeds, initiating bubble nucleation. The corresponding temperature is known as the nucleation temperature ($T_n$) which can be determined using the relation
\begin{align}
	\Gamma_d(T_n) \simeq \mathcal{H}(T_n)^4 ,  \label{Nucleation condition}
\end{align}
where, ${\mathcal{H}}(T) = 0.33 \sqrt{g_*}\frac{T^2}{M_P}$ ($M_P = 2.4 \times 10^{18}$ GeV is the reduced Planck mass) and the false-vacuum decay rate is~\cite{Coleman:1977py,Linde:1980tt,Linde:1981zj} 
\begin{align}
	\Gamma_d(T) \simeq  T^4 \left( \frac{S_3}{2\pi T} \right)^{3/2} {\rm exp}(-S_3/T) ,
\end{align}
where, $S_3$, denoting the 3-dimensional Euclidean action for $O(3)$-symmetric~\cite{Linde:1981zj} bounce solution~\cite{Coleman:1977py}, is given by
\begin{align}
	S_3 = 4\pi \int dr r^2  \left(   \frac{1}{2}  \left( \frac{dh}{d r}\right)^2 +  \frac{1}{2}  \left( \frac{dS}{d r}\right)^2  + V_{\rm eff}(h,S,T)  \right) .
\end{align}
We have used the Mathematica based package {\bf FindBounce}~\cite{Guada:2020xnz} to compute the action $S_3(T)$ numerically and finally, $T_n$ is determined using the relation 
\begin{align}
	\frac{S_3(T_n)}{T_n} \simeq 140,  \label{action-140}
\end{align}
obtained from Eq.~\ref{Nucleation condition}. In Table~\ref{tab:Two Benchmark}, we provide two sets of parameter values for $\lambda_s, \lambda_{hs}, \mu_s^2$ as benchmarks which fix $T_s, T_c$ and $T_n$, satisfying the condition of FOEWPT and the successful completion of phase transition, for the rest of our analysis. 
 \begin{table}[!htb]
	\begin{tabular}{| c | c | c | c || c| c | c |}
		\hline
		 & $\lambda_{s}$  & $\lambda_{hs}$ & $\mu_s^2$ (GeV$^2$) & $T_s$ (GeV) & $T_c$ (GeV) & $T_n$ (GeV)  \\
		\hline
		BP1 & 0.03 &  0.22  &  -2240.5  &  225.23  &  100  &  42.7   \\
		\hline
		BP2 & 0.1 &  0.37  &  -4367.5  &  224.49  &  95  &  51.4   \\
		\hline
	\end{tabular}
	\caption{Benchmark points satisfying FOEWPT}
	\label{tab:Two Benchmark}
\end{table}

We now proceed to study the QCD axion dynamics across the three phases (I$\rightarrow$III as described earlier) in the context of the standard misalignment mechanism taking into the axion-mass variation (depending upon the $vev$ structure of $S$ field) across the FOEWPT. The global PQ symmetry being broken during inflation, the axion field $a({\bf x},t)$ becomes spatially uniform, $i.e. ~a(t)$ only. Till the Hubble expansion rate {$\mathcal{H}(T)$} (considering radiation domination) exceeds the axion mass $m_{a}(T)$, it remains frozen at its initial value $a_{i}$, which is characterized by the misalignment angle $\theta_{i}$ and hence $\dot{\theta}_{i}=0$. The classical equation of motion of the axion field $a(T)$ (parametrized by $\theta(T) = a(T)/f_a$) is given by
\begin{equation}
	\ddot{\theta} + 3 \mathcal{H}(T) \dot{\theta} + m^2_{a}(T) \sin\theta =0. 
	\label{axion-evolution-eq}
\end{equation}
where ``dot" denotes the derivative with respect to time $t$. Note that, we keep the axion decay constant $f_a$ ($\geq 10^8$ GeV) and the cut-off scale $\Lambda$ (where $f_a < \Lambda < M_{Pl}$) as free parameters in studying the axion dynamics such that their values can be constrained from DM relic satisfaction. Other parameters involved in the potential $V_{\rm eff}$ are guided by BP1.\\
\vskip -0.25cm
\noindent Phase-{\bf{I}} $(T>T_s)$: The axion mass does not get any additional contribution from the higher-order PQ-symmetry breaking term. On the other hand, the lower bound of $f_a$ along with the suppression factor $(T/T_{\rm QCD})^{-b}$  makes the effective mass parameter $m_a(T)$ to be small enough not to start any oscillation in this regime. As a result, we expect the axion to remain at its initial condition. \\
\vskip -0.25cm
\begin{figure}[h]
	\includegraphics[width=1\linewidth]{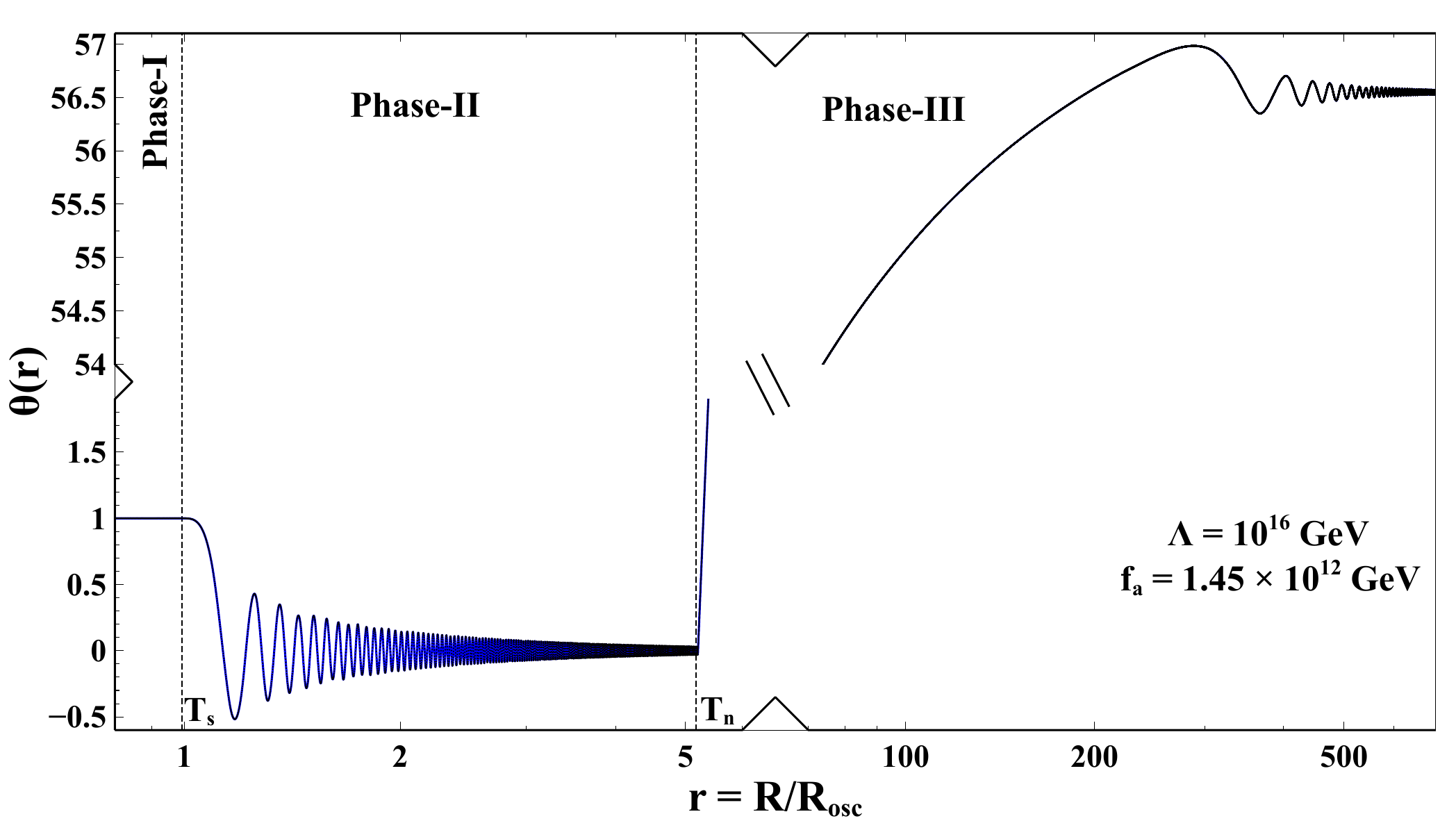}
	\caption{{\small{Evolution of $\theta$ across the three phases: before, during, and after the EWPT. Breaks in both axes are used to clearly show both early and late-time dynamics. }}}
	\label{fig:theta-r}
\end{figure}
\noindent Phase-{\bf{II}} $(T_n<T\leq T_s)$: The $S$ field gains a non zero $vev ~v_s(T)$, and the axion mass $m_a(T)$ gets modified by the $S$-portal interaction term as provided in Eq.~\ref{eq:mass-variation}. With appropriate 
choice of $v_s(T)/\Lambda$ ($v_s(T)$ is a derived parameter as per Table~\ref{tab:Two Benchmark}), such a modified axion mass can easily induce an early era of QCD axion oscillation at a temperature $T_{\rm osc}^{\rm I}$ in between $T_s$ and $T_n$, guided by $m_{a}(T_{\rm osc}^{\rm I}) \geq 3 \mathcal{H}(T_{\rm osc}^{\rm I})$ condition. Such a situation is depicted in 
Fig.~\ref{fig:theta-r}, where $R_{\rm osc}$ corresponds to the scale factor of the Universe at $T_{\rm osc}^{\rm I}$. The plot results from the numerical solution of the axion evolution equation. We note that the oscillation with decreasing amplitude continues till $m_{a}(T) = 3 \mathcal{H}(T)$ is met. \\
\vskip -0.25cm
\noindent Phase-{\bf{III}} $(T\leq T_n)$: At $T_n$, $v_s$ drops to zero, enabling the axion mass $m_a(T)$ to decrease sharply. As a result, again a situation $m_a(T) \ll \mathcal{H}(T)$ prevails and the axion oscillation ceases as it suddenly feels a strong Hubble friction. However, the axion field does not get frozen immediately in this case, rather it evolves over field space with a non-zero $\dot{\theta}$. 
This is because the sudden shift in the $vev$ of the $S$ field brings an abrupt change in the mass and the structure of the axion potential. Although this transition effectively halts the oscillatory behaviour, the axion retains a residual kinetic energy from its earlier dynamics and leaps across topologically different vacua until it loses energy to be trapped in a specific minimum. This behaviour is shown in the initial part of region-III of Fig.~\ref{fig:theta-r} with an increasing $\theta$ till a point ($T = T_{\rm osc}^{\rm II}$) where the condition $m_{a}(T_{\rm osc}^{\rm II}) = 3 \mathcal{H}(T_{\rm osc}^{\rm II})$ is again satisfied. Hence, a second oscillation phase of the QCD axion begins at $T_{\rm osc}^{\rm II}$ as shown. 

Though the beginning of this second oscillatory phase is essentially guided by the original QCD axion mass $m_{a0}$, an interesting feature emerges due to intermediate dynamics of axion between $T_s - T_n$ and $T_n - T_{\rm osc}^{\rm II}$. 
As it continued to evolve across the field space below $T_n$ due to its residual kinetic energy, by the time the second oscillation begins, the field value of the axion ($\theta_{i}^{\rm II}$) has shifted from its original post-inflationary value $\theta_i \sim \mathcal{O}(1)$. This drift leads to a modified initial condition for the resumed oscillation in terms of $\theta_{i}^{\rm II}$ and non-zero $\dot{\theta}$, which significantly impacts the final axion relic density calculation as below.


Solving the Eq.~\ref{axion-evolution-eq} at each phase with the respective initial conditions for $\theta$ and $\dot{\theta}$ and appropriate mass evolution via Eq.~\ref{eq:mass-variation} while employing the relevant $T_s, T_n$ values using the 
BP1 for the parameters of FOEWPT, we evaluate the total energy density of the QCD axion at $T_{\rm QCD}$, given by
\begin{equation}
	\rho_a\left(T_{\rm QCD}\right) =\left[ \frac{1}{2} f_a^2 \dot{\theta}^2 + m_a^2(T) f_a^2 \left( 1 - \cos\theta \right) \right]_{T=T_{\rm QCD}}.
\end{equation}
Since the number density of the axion in a comoving volume remains conserved, the axion energy density at the present day can be estimated as
\begin{equation}
	\rho_0\left(T_0\right) = \rho_a\left(T_{\rm QCD}\right) \left( \frac{R_{\rm QCD}}{R_0} \right)^3 \frac{m_a\left(T_0\right)}{m_a\left(T_{\rm QCD}\right)}, \label{eq:axion-energy-present-day}
\end{equation}
where, $T_0=2.4 \times 10^{-4}$ eV is the present temperature. 
\begin{figure}[!htb]
	\includegraphics[width=1\linewidth]{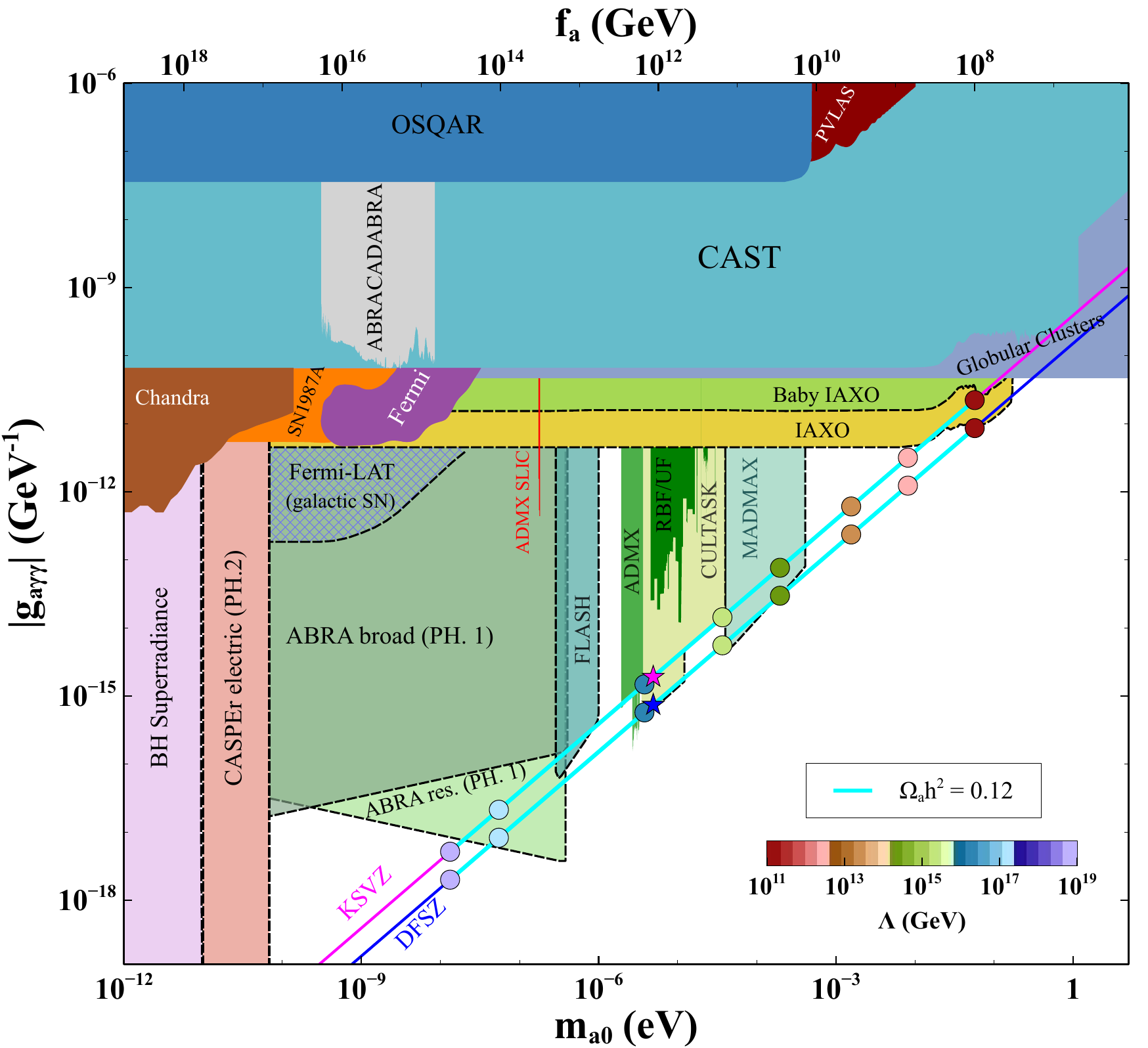}
	\caption{{{Sensitivities of experimental and observational constraints in the axion-photon coupling $(g_{a\gamma\gamma})$ - $m_{a0}$ ($f_a$) plane with relic satisfied points in our case. The projected sensitivities are marked with dashed lines. All the bounds are taken from the online repository \texttt{AxionLimits} \cite{Ohare2020-gy}.}}}
	\label{fig:ma-axion-photon}
	\end{figure}
Here we consider the expansion of the Universe as adiabatic in nature satisfying $\left(R_{\rm QCD}/R_0\right)^3 = s(T_0)/s(T_{\rm QCD})$, where $s(T)$ is the entropy density of the Universe at $T$ given by $s(T)= (2\pi^2/45) g_{*s} T^3$. The axion relic density would then be 
\begin{equation}
	\Omega_a h^2 =\frac{h^2}{\rho_{c,0}} \rho_a(T_{\rm QCD}) \left[\frac{(m_a\, g_{*s})_{T_0}}{(m_a\, g_{*s})_{T_{\rm QCD}}}\right]\left( \frac{T_0}{T_{\rm QCD}} \right)^3,
\label{eq:Axion-relic-density}
\end{equation}
where, $\rho_{c,0} = 1.05 \times 10^{-5}\,h^2$ GeV ${\rm cm^{-3}}$ denotes the present critical energy density and $g_{*s}(T_0)=3.94$ is the effective number of relativistic d.o.f at the present temperature $T_0$. We then study the axion dynamics for several $f_a$ values to look for their relic satisfaction by varying $\Lambda$. We find the entire range of $f_a$ as $10^{8-14}$ GeV is allowed for the QCD axion being dark matter. The upper limit on $f_a$ follows from the requirement of the cut-off scale $\Lambda$ to remain below the Planck scale. 
\begin{figure}[!htb]
	\includegraphics[width=1\linewidth]{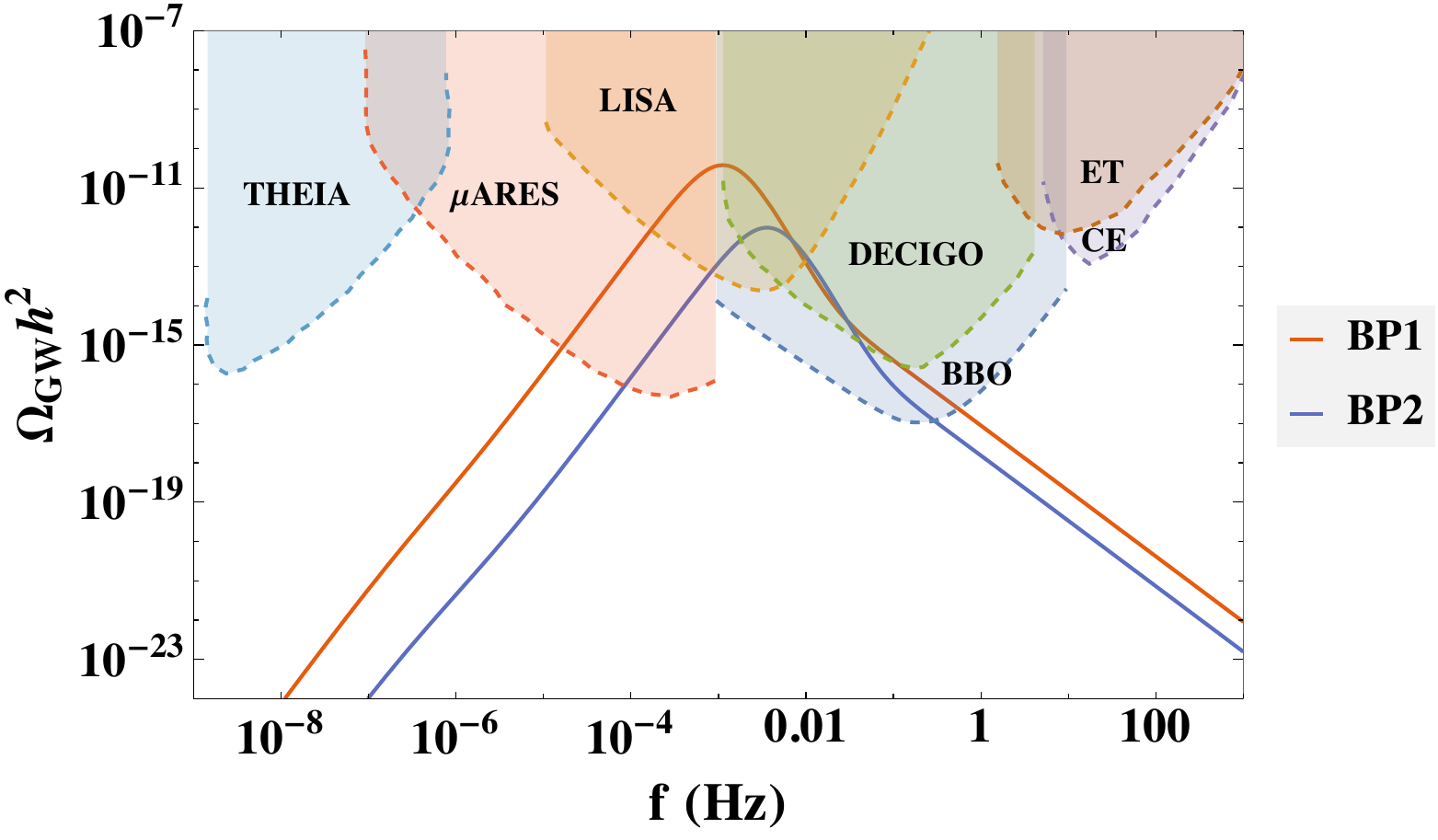}
	\caption{{\small{GW spectrum for FOEWPT}}}
	\label{fig:GW}
\end{figure}
To make it explicit, few such $m_{a0} (f_a) - \Lambda$ sets are presented in Fig.~\ref{fig:ma-axion-photon} by coloured (colour code for $\Lambda$ is indicated inset) circles on the axion line(s) in axion-photon coupling {($g_{a\gamma\gamma} $)} versus $m_{a0} (f_a)$ plane.    
The axion-photon coupling $g_{a\gamma\gamma} $ is defined as~\cite{GrillidiCortona:2015jxo}:
\begin{equation}
	g_{a\gamma\gamma} = \frac{\alpha}{2\pi f_a} \left( \frac{ E}{N} - 1.92 \right).
\end{equation}
$E$ and $N$ are defined as the electromagnetic and color anomalies of the axial current associated with the axion. For DFSZ~\cite{DINE1981199,Zhitnitsky:1980tq}, $E/N =8/3$ and for KSVZ~\cite{Kim:1979if,Shifman:1979if} model $E/N = 0 $.
This scenario substantially enlarges the viable parameter space for the QCD axion as a DM candidate and enhances its detection prospects in upcoming experiments such as ABRACADABRA \cite{Kahn:2016aff,Ouellet:2018beu}, KLASH \cite{Alesini:2017ifp,Alesini:2019nzq} and FLASH \cite{Alesini:2023qed}, CULTASK \cite{Lee:2020cfj,Semertzidis:2019gkj}, IAXO \cite{Vogel:2013bta,IAXO:2019mpb} and MADMAX \cite{Caldwell:2016dcw}. Alongside, we estimate the GW spectrum, shown in Fig.~\ref{fig:GW}, for the two benchmark points listed in Table~\ref{tab:Two Benchmark} associated with the FOEWPT in our scenario as a possible probe within the sensitivity reach of various proposed GW detectors such as LISA \cite{LISA:2017pwj}, BBO \cite{Yagi:2011wg,Crowder:2005nr,Corbin:2005ny,Harry:2006fi}, DECIGO \cite{Yagi:2011wg,Kawamura:2006up}, $\mu$ARES \cite{Sesana:2019vho}, CE \cite{LIGOScientific:2016wof,Reitze:2019iox}, ET \cite{Punturo:2010zz,Hild:2010id,Sathyaprakash:2012jk,ET:2019dnz} and THEIA \cite{Garcia-Bellido:2021zgu} (see appendix~\ref{ap:GW-Spectrum} for the detailed formulae used in the computation).

The EWPT in the early Universe is an important event, the nature of which is still unknown to us. On the other hand, the dynamics of QCD axion at high temperature is also obscure apart from the conjecture that it would start oscillating around 
its mass equivalent temperature. In this present work, we ask the question that whether there could have been an interesting interplay in the early Universe between the EWPT (while first order) and the mass (hence dynamics) of the axion field. We find that the proposed {\it{recurrent ~misalignment}} mechanism has the potential to allow the decay constant $f_a$ in a broad range which was otherwise not allowed by the relic satisfaction. The intermittent modification of the axion mass, via the non-renormalizable PQ breaking term, plays a pivotal role in realizing this. This finding is not only compelling from the detection prospects of QCD axion as DM by upcoming experiments within this extended parameter space, but also can open up other interesting phenomenological studies. In particular, since a FOEWPT is intricately connected to electroweak baryogenesis, exploring the QCD axion as a DM candidate within this framework is phenomenologically appealing. Furthermore, the same scenario can also be studied in the context of axion like particles on which we are working on. In summary, this work presents a testable framework linking axion DM with FOEWPT, which is further associated with a stochastic gravitational wave background potentially observable in near-future detectors.

\vskip 1 cm

\begin{acknowledgements}
	The work of DB is supported by Council of Scientific \& Industrial Research (CSIR), Govt. of India, under the senior research fellowship scheme. The work of AS is supported during its initial stage by the grants CRG/2021/005080 and MTR/2021/000774 from SERB, Govt. of India. AS also acknowledges the hospitality of CERN theory division during a visit where part of this work was formulated.
\end{acknowledgements}

\appendix
\onecolumngrid
\section{Origin of our dimension-6 operator} \label{ap:UV-completion}

It is pertinent here to discuss a possible origin of our choice of $U(1)$ symmetry breaking dimension-6 operator, $\frac{S^4}{\Lambda^2} \left( \Phi^2 + \Phi^{*2}\right)  e^{i\alpha}$. To do so, we will employ the fact that any global symmetry is explicitly broken by gravity, resulting Planck-suppressed higher dimensional explicit-breaking terms. With this in mind, we consider the following Lagrangian as
\beeq
-\mathcal{L}\supset \mu \Phi^2 e^{i\alpha}\psi + \frac{S^4 \psi^\dagger}{M_{P}} +h.c..
\label{appen-eq1}
\eeq
Here, $\psi$ denotes a heavy complex SM singlet scalar field with a $U(1)$ charge $-2$ and also, an even $Z_2$ charge, whereas, $\Phi$ is the $U(1)$ breaking scalar of our consideration with $U(1)$ charge +1. Therefore, the Lagrangian in Eq. \ref{appen-eq1} is $Z_2$ invariant. However, as evident from Eq. \ref{appen-eq1}, while the first term is $U(1)$ conserving (with $\mu$ being a dimensionful coupling of order $M_P$), the second term is a dimension-5 explicit $U(1)$ breaking term suppressed by $M_P$, which can emerge due to the quantum gravity effect. Assuming the hierarchy, $M_\psi < \mu~(\simeq M_P)$, where $M_\psi$ denotes the mass of the heavy scalar field $\psi$, one can integrate out $\psi$ and obtain the following dimension-6 operator at energies below $M_\psi$:
\begin{equation}
\frac{S^4}{M_\psi^2}\,  \Phi^2 \, e^{i\alpha} + h.c..
\end{equation}
Rewriting $M_\psi$ as $\Lambda$, we can easily obtain the dimension-6 operator of our interest. The main advantage of such construction lies in the fact that it forbids other possible $U(1)$-breaking dimension-6 operators, constructed out of the scalar fields (e.g., $|H|^4\left( \Phi^2 + \Phi^{*2}\right)$, $|\Phi|^4 \left( \Phi^2 + \Phi^{*2}\right)$, $|H|^2 S^2 \left( \Phi^2 + \Phi^{*2}\right)$, $|H|^2 |\Phi|^2 \left( \Phi^2 + \Phi^{*2}\right)$ etc.), from appearing in the picture. 

\section{Effective Potential construction} \label{ap:effective-potential}
We extend the SM Higgs sector by a $Z_2$ symmetric real singlet scalar $S$ which couples with the Higgs generating a tree-level barrier, thereby setting the platform for first-order EWPT. 

The tree-level potential, corresponds to the classical background fields $h$ and $S$, can be written as
\begin{align}
	V_0(h,S) =\, &  \frac{1}{2} \mu_h^2  h^2 + \frac{\lambda_h}{4} h^4 +  \frac{1}{2}  \mu_s^2  S^2 + \frac{\lambda_s}{4} S^4  +  \frac{\lambda_{hs}}{4} h^2 S^2 .  
\end{align}

To study the phase transition in the early Universe, we have to consider the one-loop zero-temperature correction as well as one-loop thermal contribution to the tree-level potential along with the corrections due to ring diagrams at higher loop, called daisy resummation~\cite{Arnold:1992rz,Carrington:1991hz}. Thereby, the full effective potential can be written as
\begin{align}
	V_{\rm eff}(h,S,T) = V_{0}(h,S) + V_{\rm CW}(h,S) & + V_{\rm T}(h,S,T) +  V_{\rm daisy}(h,S,T) , \label{ap-effective-potential}
\end{align}
The one-loop thermal correction~\cite{Quiros:1999jp} is given by
\begin{equation}
	V_{T}(h,S,T)=\frac{T^4}{2\pi^2} \left[\sum_{i=h,S,G, W,Z}n_i J_B\left(\frac{m_i^2(h,S)}{T^2}\right)-\sum_{i=t} n_i J_F\left(\frac{m_i^2(h)}{T^2}\right) \right], \label{thermal-potential-1}
\end{equation}
where the thermal functions $J_{B,F}$ are defined as
\begin{equation}
	J_{B,F}\left(\frac{m_i^2}{T^2}\right)=\int_0^{\infty} dx x^2 \log\left[1\mp {\rm exp}\left(-\sqrt{\frac{x^2+m_i^2}{T^2}}\right)\right]. \label{thermal-integral-1}
\end{equation}
Here $\mp$ signs are for bosons and fermions respectively. The factors 
\begin{align}
	n_h=1, \,\, n_s=1, \,\,  n_G=3, \,\, n_W=6, \,\,  n_Z=3,  \,\,  n_t=12 ,
\end{align} 
represent the degrees of freedom of the Higgs, the $S$ field, the Goldstone bosons, $W$, $Z$ bosons and the top quark respectively. The field dependent masses denoted by $m_i$ are given by:
\begin{align}
	m_h^2(h,S) & = \mu_h^2 +3 \lambda_h h^2 + \frac{\lambda_{hs}}{2} S^2  \nonumber \\ 
	m_G^2(h,S) & =  \mu_h^2 + \lambda_h h^2 + \frac{\lambda_{hs}}{2} S^2   \nonumber \\
	m_s^2(h,S) & = \mu_s^2 + 3 \lambda_s S^2 + 	\frac{\lambda_{hs}}{2} h^2  \nonumber \\
	m_{W}^2(h) & =  \frac{g_{\rm w}^2}{4} h^2, \quad 
	m_{Z}^2(h) = \frac{g_{\rm w}^2 + g_{Y}^2}{4} h^2, \quad  
	m_t^2(h)  = \frac{y_t^2}{2} h^2,  \nonumber
\end{align}
where $g_{\rm w}$ and $g_Y$ are the gauge coupling strengths and $y_t$ denotes the top quark yukawa coupling. \\
We neglect the one-loop Coleman-Weinberg (zero-temperature) contribution~\cite{Coleman:1973jx} and construct the effective potential using the high-temperature expansion for the thermal corrections.

In the high-temperature expansion, $J_{B,F}(m^2/T^2)$ take the form as
\begin{equation}
	J_B(m^2/T^2)  = - \frac{\pi^4}{45} +  \frac{\pi^2}{12}  \frac{m^2}{T^2} -  \frac{\pi}{6}  \left(\frac{m^2}{T^2}\right)^{3/2} + ... \hspace{0.1 cm}, \quad J_F(m^2/T^2) = -  \frac{7\pi^4}{360} -  \frac{\pi^2}{24}  \frac{m^2}{T^2} + ...  \hspace{0.1 cm} .
\end{equation}
Taking into account terms upto order $T^2$ and hence ignoring the daisy corrections, the effective potential Eq.~(\ref{ap-effective-potential}) can be written as: 
\begin{align}
	V_{\rm eff}(h,S,T) =  \frac{1}{2}( \mu_h^2 + c_h T^2) h^2 + \frac{\lambda_h}{4} h^4 + \frac{1}{2}( \mu_s^2 + c_s T^2) S^2 + \frac{\lambda_s}{4} S^4 +  \frac{\lambda_{hs}}{4} h^2 S^2 , \label{ap-effective-high-T-potential}
\end{align}
which represents the effective potential for Higgs and $S$ field as we considered in the main manuscript for studying the electroweak phase transition. $c_h$ and $c_s$ are defined in Eq.~\ref{ch-cs} in the main manuscript.

\section{Calculation of Gravitational Wave Spectrum} \label{ap:GW-Spectrum}
In a cosmological first-order phase transition, gravitational waves (GW) are primarily generated through three mechanisms: bubble collisions, sound waves in the plasma, and magnetohydrodynamic (MHD) turbulence. The relative importance of each contribution is governed by the dynamics of the transition, especially the velocity of the expanding bubble walls, denoted by $v_w$.

This work focuses on a scenario in which the bubble wall propagates subsonically through the plasma, i.e., with a velocity $v_w<c_s$, where $c_s=1/\sqrt{3}$ being the sound speed in the plasma. In such a non-runaway bubble wall regime, the gravitational wave spectrum is predominantly sourced by sound waves and MHD turbulence within the plasma. The total GW spectrum can be expressed as~\cite{Caprini:2015zlo}
\begin{equation}
	\Omega_{\rm GW}h^2 \simeq \Omega_{\rm sw} h^2 + \Omega_{\rm turb} h^2.
\end{equation}
The contributions from the sound waves and MHD turbulence are, respectively, given by~\cite{Athron:2023xlk,Guo:2020grp,Caprini:2019egz,Hindmarsh:2020hop}
\begin{align}
	\Omega_{\rm sw}(f) h^2 & =\frac{1.23 \times 10^{-5}}{g_*^{1/3}} \frac{H_*}{\beta} \left( \frac{k_{\rm sw} \alpha_*}{1+\alpha_*} \right)^2 v_w S_{\rm sw}(f) \Upsilon , \\
	\Omega_{\rm turb}(f) h^2 & =\frac{1.55 \times 10^{-3}}{g_*^{1/3}} \frac{H_*}{\beta} \left( \frac{k_{\rm turb} \alpha_*}{1+\alpha_*} \right)^{\frac{3}{2}} v_w S_{\rm turb}(f).
\end{align}
Here, $\Upsilon = 1 - \frac{1}{\sqrt{1 + 2 \tau_{\rm sw}  H_*}}$ is a suppression factor~\cite{Guo:2020grp} where $\tau_{\rm sw} \sim R_*/\bar{U_f} $ with $R_* = (8\pi)^{1/3} v_w \beta^{-1}$ be the mean bubble separation and $\bar{U_f}= \sqrt{\frac{3 k_{\rm sw} \alpha_*}{4(1 + \alpha_*)}} $ be the rms fluid velocity.\\
The fraction $\beta/H_*$, with $\beta$ being the inverse time duration of the phase transition
and $H_*$ being the Hubble constant at temperature $T_n$, can be evaluated as~\cite{Grojean:2006bp}
\begin{equation}
	\frac{\beta}{H_*} \simeq T_n \left. \frac{d}{dT}\left( \frac{S_3}{T} \right) \right|_{T=T_n}
\end{equation}
assuming $T_n$ is the temperature of the plasma when GW is produced.\\
The quantity $\alpha_*$, denoting the ratio of released vacuum energy in the phase transition to that of the radiation bath at $T=T_n$, can be written as~\cite{Grojean:2006bp}
\begin{equation}
	\alpha_* = \frac{1}{\rho^*_{\rm rad}} \left[ \Delta V - \frac{T}{4} \frac{\partial \Delta V}{\partial T} \right]_{T=T_n},
\end{equation}
where $\Delta V = V_{\rm eff}(0,v_s(T),T)-V_{\rm eff}(v_h(T),0,T)$ and $\rho^*_{\rm rad} = g_* \pi^2 T_n^4/30$ is the radiation energy density at $T_n$.

The functions parametrizing the spectral shape of the GWs read as~\cite{Caprini:2015zlo,Vaskonen:2016yiu}
\begin{align}
	S_{\rm sw}(f) = \left( \frac{f}{f_{\rm sw}} \right)^3 \left( \frac{7}{4+3(f/f_{\rm sw})^2}  \right)^{7/2} ,\quad
	S_{\rm turb}(f) = \frac{(f/f_{\rm turb})^3}{(1+ (f/f_{\rm trub}))^{\frac{11}{3}} (1+ 8\pi f/h_*)},
\end{align}
with 
\begin{equation}
	h_*=1.65 \times 10^{-5} \hspace{0.04 cm}{\rm Hz} \left( \frac{T_n}{100 \hspace{0.04 cm} {\rm GeV}} \right)  \left( \frac{g_*}{100 } \right)^{\frac{1}{6}}.
\end{equation}
Here $f_{\rm sw}$ and $f_{\rm turb}$ are the peak frequencies of each contribution can be written as
\begin{align}
	f_{\rm sw}  = \frac{1.9 \times 10^{-5} \hspace{0.04 cm}{\rm Hz}}{v_w} \frac{\beta}{H_*} \left( \frac{T_n}{100 \hspace{0.04 cm} {\rm GeV}} \right)  \left( \frac{g_*}{100 } \right)^{\frac{1}{6}} , \quad
	f_{\rm turb} = 1.42 f_{\rm sw} .
\end{align}
$k_{\rm sw}$ and $k_{\rm turb}$ are the fractions of the released vacuum energy density converted into bulk motion of fluid and MHD turbulence, respectively. For subsonic bubble walls these can be defined as~\cite{Espinosa:2010hh}
\begin{align}
	k_{\rm sw}  = \frac{c_s^{11/5} k_{\rm a}k_{\rm b}}{(c_s^{11/5}-v_w^{11/5})k_{\rm b} + v_w c_s^{6/5} k_{\rm a}} , \quad
	k_{\rm turb}  = \epsilon k_{\rm sw} ,
\end{align}
with $\epsilon$ typically in the range of $5\%$-$10\%$~\cite{Caprini:2015zlo,Hindmarsh:2015qta}. We use $\epsilon=0.05$ for conservative choice. $k_{\rm a}$ and $k_{\rm b}$ are defined as~\cite{Espinosa:2010hh}
\begin{align}
	k_{\rm a} = \frac{6.9 v_w^{6/5}\alpha_*}{1.36 - 0.037 \sqrt{\alpha_*} + \alpha_*} ,\quad
	k_{\rm b} = \frac{\alpha_*^{2/5}}{0.017 + (0.997 + \alpha_*)^{2/5}}.
\end{align}

\twocolumngrid
\bibliography{Ref-1.bib}

\begin{thebibliography}{99}%
\makeatletter
\providecommand \@ifxundefined [1]{%
 \@ifx{#1\undefined}
}%
\providecommand \@ifnum [1]{%
 \ifnum #1\expandafter \@firstoftwo
 \else \expandafter \@secondoftwo
 \fi
}%
\providecommand \@ifx [1]{%
 \ifx #1\expandafter \@firstoftwo
 \else \expandafter \@secondoftwo
 \fi
}%
\providecommand \natexlab [1]{#1}%
\providecommand \enquote  [1]{``#1''}%
\providecommand \bibnamefont  [1]{#1}%
\providecommand \bibfnamefont [1]{#1}%
\providecommand \citenamefont [1]{#1}%
\providecommand \href@noop [0]{\@secondoftwo}%
\providecommand \href [0]{\begingroup \@sanitize@url \@href}%
\providecommand \@href[1]{\@@startlink{#1}\@@href}%
\providecommand \@@href[1]{\endgroup#1\@@endlink}%
\providecommand \@sanitize@url [0]{\catcode `\\12\catcode `\$12\catcode
  `\&12\catcode `\#12\catcode `\^12\catcode `\_12\catcode `\%12\relax}%
\providecommand \@@startlink[1]{}%
\providecommand \@@endlink[0]{}%
\providecommand \url  [0]{\begingroup\@sanitize@url \@url }%
\providecommand \@url [1]{\endgroup\@href {#1}{\urlprefix }}%
\providecommand \urlprefix  [0]{URL }%
\providecommand \Eprint [0]{\href }%
\providecommand \doibase [0]{https://doi.org/}%
\providecommand \selectlanguage [0]{\@gobble}%
\providecommand \bibinfo  [0]{\@secondoftwo}%
\providecommand \bibfield  [0]{\@secondoftwo}%
\providecommand \translation [1]{[#1]}%
\providecommand \BibitemOpen [0]{}%
\providecommand \bibitemStop [0]{}%
\providecommand \bibitemNoStop [0]{.\EOS\space}%
\providecommand \EOS [0]{\spacefactor3000\relax}%
\providecommand \BibitemShut  [1]{\csname bibitem#1\endcsname}%
\let\auto@bib@innerbib\@empty
\bibitem [{\citenamefont {Peccei}\ and\ \citenamefont
  {Quinn}(1977{\natexlab{a}})}]{Peccei:1977hh}%
  \BibitemOpen
  \bibfield  {author} {\bibinfo {author} {\bibfnamefont {R.~D.}\ \bibnamefont
  {Peccei}}\ and\ \bibinfo {author} {\bibfnamefont {H.~R.}\ \bibnamefont
  {Quinn}},\ }\bibfield  {title} {\bibinfo {title} {{CP Conservation in the
  Presence of Instantons}},\ }\href
  {https://doi.org/10.1103/PhysRevLett.38.1440} {\bibfield  {journal} {\bibinfo
   {journal} {Phys. Rev. Lett.}\ }\textbf {\bibinfo {volume} {38}},\ \bibinfo
  {pages} {1440} (\bibinfo {year} {1977}{\natexlab{a}})}\BibitemShut {NoStop}%
\bibitem [{\citenamefont {Peccei}\ and\ \citenamefont
  {Quinn}(1977{\natexlab{b}})}]{Peccei:1977ur}%
  \BibitemOpen
  \bibfield  {author} {\bibinfo {author} {\bibfnamefont {R.~D.}\ \bibnamefont
  {Peccei}}\ and\ \bibinfo {author} {\bibfnamefont {H.~R.}\ \bibnamefont
  {Quinn}},\ }\bibfield  {title} {\bibinfo {title} {{Constraints Imposed by CP
  Conservation in the Presence of Instantons}},\ }\href
  {https://doi.org/10.1103/PhysRevD.16.1791} {\bibfield  {journal} {\bibinfo
  {journal} {Phys. Rev. D}\ }\textbf {\bibinfo {volume} {16}},\ \bibinfo
  {pages} {1791} (\bibinfo {year} {1977}{\natexlab{b}})}\BibitemShut {NoStop}%
\bibitem [{\citenamefont {Weinberg}(1978)}]{Weinberg:1977ma}%
  \BibitemOpen
  \bibfield  {author} {\bibinfo {author} {\bibfnamefont {S.}~\bibnamefont
  {Weinberg}},\ }\bibfield  {title} {\bibinfo {title} {{A New Light Boson?}},\
  }\href {https://doi.org/10.1103/PhysRevLett.40.223} {\bibfield  {journal}
  {\bibinfo  {journal} {Phys. Rev. Lett.}\ }\textbf {\bibinfo {volume} {40}},\
  \bibinfo {pages} {223} (\bibinfo {year} {1978})}\BibitemShut {NoStop}%
\bibitem [{\citenamefont {Wilczek}(1978)}]{Wilczek:1977pj}%
  \BibitemOpen
  \bibfield  {author} {\bibinfo {author} {\bibfnamefont {F.}~\bibnamefont
  {Wilczek}},\ }\bibfield  {title} {\bibinfo {title} {{Problem of Strong $P$
  and $T$ Invariance in the Presence of Instantons}},\ }\href
  {https://doi.org/10.1103/PhysRevLett.40.279} {\bibfield  {journal} {\bibinfo
  {journal} {Phys. Rev. Lett.}\ }\textbf {\bibinfo {volume} {40}},\ \bibinfo
  {pages} {279} (\bibinfo {year} {1978})}\BibitemShut {NoStop}%
\bibitem [{\citenamefont {Crewther}\ \emph {et~al.}(1979)\citenamefont
  {Crewther}, \citenamefont {Di~Vecchia}, \citenamefont {Veneziano},\ and\
  \citenamefont {Witten}}]{Crewther:1979pi}%
  \BibitemOpen
  \bibfield  {author} {\bibinfo {author} {\bibfnamefont {R.~J.}\ \bibnamefont
  {Crewther}}, \bibinfo {author} {\bibfnamefont {P.}~\bibnamefont
  {Di~Vecchia}}, \bibinfo {author} {\bibfnamefont {G.}~\bibnamefont
  {Veneziano}},\ and\ \bibinfo {author} {\bibfnamefont {E.}~\bibnamefont
  {Witten}},\ }\bibfield  {title} {\bibinfo {title} {{Chiral Estimate of the
  Electric Dipole Moment of the Neutron in Quantum Chromodynamics}},\ }\href
  {https://doi.org/10.1016/0370-2693(79)90128-X} {\bibfield  {journal}
  {\bibinfo  {journal} {Phys. Lett. B}\ }\textbf {\bibinfo {volume} {88}},\
  \bibinfo {pages} {123} (\bibinfo {year} {1979})},\ \bibinfo {note} {[Erratum:
  Phys.Lett.B 91, 487 (1980)]}\BibitemShut {NoStop}%
\bibitem [{\citenamefont {Abel}\ \emph {et~al.}(2020)\citenamefont {Abel} \emph
  {et~al.}}]{Abel:2020pzs}%
  \BibitemOpen
  \bibfield  {author} {\bibinfo {author} {\bibfnamefont {C.}~\bibnamefont
  {Abel}} \emph {et~al.},\ }\bibfield  {title} {\bibinfo {title} {{Measurement
  of the Permanent Electric Dipole Moment of the Neutron}},\ }\href
  {https://doi.org/10.1103/PhysRevLett.124.081803} {\bibfield  {journal}
  {\bibinfo  {journal} {Phys. Rev. Lett.}\ }\textbf {\bibinfo {volume} {124}},\
  \bibinfo {pages} {081803} (\bibinfo {year} {2020})},\ \Eprint
  {https://arxiv.org/abs/2001.11966} {arXiv:2001.11966 [hep-ex]} \BibitemShut
  {NoStop}%
\bibitem [{\citenamefont {Preskill}\ \emph {et~al.}(1983)\citenamefont
  {Preskill}, \citenamefont {Wise},\ and\ \citenamefont
  {Wilczek}}]{Preskill:1982cy}%
  \BibitemOpen
  \bibfield  {author} {\bibinfo {author} {\bibfnamefont {J.}~\bibnamefont
  {Preskill}}, \bibinfo {author} {\bibfnamefont {M.~B.}\ \bibnamefont {Wise}},\
  and\ \bibinfo {author} {\bibfnamefont {F.}~\bibnamefont {Wilczek}},\
  }\bibfield  {title} {\bibinfo {title} {{Cosmology of the Invisible Axion}},\
  }\href {https://doi.org/10.1016/0370-2693(83)90637-8} {\bibfield  {journal}
  {\bibinfo  {journal} {Phys. Lett. B}\ }\textbf {\bibinfo {volume} {120}},\
  \bibinfo {pages} {127} (\bibinfo {year} {1983})}\BibitemShut {NoStop}%
\bibitem [{\citenamefont {Abbott}\ and\ \citenamefont
  {Sikivie}(1983)}]{Abbott:1982af}%
  \BibitemOpen
  \bibfield  {author} {\bibinfo {author} {\bibfnamefont {L.~F.}\ \bibnamefont
  {Abbott}}\ and\ \bibinfo {author} {\bibfnamefont {P.}~\bibnamefont
  {Sikivie}},\ }\bibfield  {title} {\bibinfo {title} {{A Cosmological Bound on
  the Invisible Axion}},\ }\href {https://doi.org/10.1016/0370-2693(83)90638-X}
  {\bibfield  {journal} {\bibinfo  {journal} {Phys. Lett. B}\ }\textbf
  {\bibinfo {volume} {120}},\ \bibinfo {pages} {133} (\bibinfo {year}
  {1983})}\BibitemShut {NoStop}%
\bibitem [{\citenamefont {Dine}\ and\ \citenamefont
  {Fischler}(1983)}]{Dine:1982ah}%
  \BibitemOpen
  \bibfield  {author} {\bibinfo {author} {\bibfnamefont {M.}~\bibnamefont
  {Dine}}\ and\ \bibinfo {author} {\bibfnamefont {W.}~\bibnamefont
  {Fischler}},\ }\bibfield  {title} {\bibinfo {title} {{The Not So Harmless
  Axion}},\ }\href {https://doi.org/10.1016/0370-2693(83)90639-1} {\bibfield
  {journal} {\bibinfo  {journal} {Phys. Lett. B}\ }\textbf {\bibinfo {volume}
  {120}},\ \bibinfo {pages} {137} (\bibinfo {year} {1983})}\BibitemShut
  {NoStop}%
\bibitem [{\citenamefont {Turner}(1983)}]{Turner:1983he}%
  \BibitemOpen
  \bibfield  {author} {\bibinfo {author} {\bibfnamefont {M.~S.}\ \bibnamefont
  {Turner}},\ }\bibfield  {title} {\bibinfo {title} {{Coherent Scalar Field
  Oscillations in an Expanding Universe}},\ }\href
  {https://doi.org/10.1103/PhysRevD.28.1243} {\bibfield  {journal} {\bibinfo
  {journal} {Phys. Rev. D}\ }\textbf {\bibinfo {volume} {28}},\ \bibinfo
  {pages} {1243} (\bibinfo {year} {1983})}\BibitemShut {NoStop}%
\bibitem [{\citenamefont {Arias}\ \emph {et~al.}(2012)\citenamefont {Arias},
  \citenamefont {Cadamuro}, \citenamefont {Goodsell}, \citenamefont {Jaeckel},
  \citenamefont {Redondo},\ and\ \citenamefont {Ringwald}}]{Arias:2012az}%
  \BibitemOpen
  \bibfield  {author} {\bibinfo {author} {\bibfnamefont {P.}~\bibnamefont
  {Arias}}, \bibinfo {author} {\bibfnamefont {D.}~\bibnamefont {Cadamuro}},
  \bibinfo {author} {\bibfnamefont {M.}~\bibnamefont {Goodsell}}, \bibinfo
  {author} {\bibfnamefont {J.}~\bibnamefont {Jaeckel}}, \bibinfo {author}
  {\bibfnamefont {J.}~\bibnamefont {Redondo}},\ and\ \bibinfo {author}
  {\bibfnamefont {A.}~\bibnamefont {Ringwald}},\ }\bibfield  {title} {\bibinfo
  {title} {{WISPy Cold Dark Matter}},\ }\href
  {https://doi.org/10.1088/1475-7516/2012/06/013} {\bibfield  {journal}
  {\bibinfo  {journal} {JCAP}\ }\textbf {\bibinfo {volume} {06}},\ \bibinfo
  {pages} {013}},\ \Eprint {https://arxiv.org/abs/1201.5902} {arXiv:1201.5902
  [hep-ph]} \BibitemShut {NoStop}%
\bibitem [{\citenamefont {Raffelt}(2008)}]{Raffelt:2006cw}%
  \BibitemOpen
  \bibfield  {author} {\bibinfo {author} {\bibfnamefont {G.~G.}\ \bibnamefont
  {Raffelt}},\ }\bibfield  {title} {\bibinfo {title} {{Astrophysical axion
  bounds}},\ }\href {https://doi.org/10.1007/978-3-540-73518-2_3} {\bibfield
  {journal} {\bibinfo  {journal} {Lect. Notes Phys.}\ }\textbf {\bibinfo
  {volume} {741}},\ \bibinfo {pages} {51} (\bibinfo {year} {2008})},\ \Eprint
  {https://arxiv.org/abs/hep-ph/0611350} {arXiv:hep-ph/0611350} \BibitemShut
  {NoStop}%
\bibitem [{\citenamefont {Caputo}\ and\ \citenamefont
  {Raffelt}(2024)}]{Caputo:2024oqc}%
  \BibitemOpen
  \bibfield  {author} {\bibinfo {author} {\bibfnamefont {A.}~\bibnamefont
  {Caputo}}\ and\ \bibinfo {author} {\bibfnamefont {G.}~\bibnamefont
  {Raffelt}},\ }\bibfield  {title} {\bibinfo {title} {{Astrophysical Axion
  Bounds: The 2024 Edition}},\ }\href {https://doi.org/10.22323/1.454.0041}
  {\bibfield  {journal} {\bibinfo  {journal} {PoS}\ }\textbf {\bibinfo {volume}
  {COSMICWISPers}},\ \bibinfo {pages} {041} (\bibinfo {year} {2024})},\ \Eprint
  {https://arxiv.org/abs/2401.13728} {arXiv:2401.13728 [hep-ph]} \BibitemShut
  {NoStop}%
\bibitem [{\citenamefont {Co}\ \emph {et~al.}(2019)\citenamefont {Co},
  \citenamefont {Gonzalez},\ and\ \citenamefont {Harigaya}}]{Co:2018phi}%
  \BibitemOpen
  \bibfield  {author} {\bibinfo {author} {\bibfnamefont {R.~T.}\ \bibnamefont
  {Co}}, \bibinfo {author} {\bibfnamefont {E.}~\bibnamefont {Gonzalez}},\ and\
  \bibinfo {author} {\bibfnamefont {K.}~\bibnamefont {Harigaya}},\ }\bibfield
  {title} {\bibinfo {title} {{Axion Misalignment Driven to the Bottom}},\
  }\href {https://doi.org/10.1007/JHEP05(2019)162} {\bibfield  {journal}
  {\bibinfo  {journal} {JHEP}\ }\textbf {\bibinfo {volume} {05}},\ \bibinfo
  {pages} {162}},\ \Eprint {https://arxiv.org/abs/1812.11186} {arXiv:1812.11186
  [hep-ph]} \BibitemShut {NoStop}%
\bibitem [{\citenamefont {Heurtier}\ \emph {et~al.}(2021)\citenamefont
  {Heurtier}, \citenamefont {Huang},\ and\ \citenamefont
  {Tait}}]{Heurtier:2021rko}%
  \BibitemOpen
  \bibfield  {author} {\bibinfo {author} {\bibfnamefont {L.}~\bibnamefont
  {Heurtier}}, \bibinfo {author} {\bibfnamefont {F.}~\bibnamefont {Huang}},\
  and\ \bibinfo {author} {\bibfnamefont {T.~M.~P.}\ \bibnamefont {Tait}},\
  }\bibfield  {title} {\bibinfo {title} {{Resurrecting low-mass axion dark
  matter via a dynamical QCD scale}},\ }\href
  {https://doi.org/10.1007/JHEP12(2021)216} {\bibfield  {journal} {\bibinfo
  {journal} {JHEP}\ }\textbf {\bibinfo {volume} {12}},\ \bibinfo {pages}
  {216}},\ \Eprint {https://arxiv.org/abs/2104.13390} {arXiv:2104.13390
  [hep-ph]} \BibitemShut {NoStop}%
\bibitem [{\citenamefont {Co}\ \emph {et~al.}(2020{\natexlab{a}})\citenamefont
  {Co}, \citenamefont {Hall},\ and\ \citenamefont {Harigaya}}]{Co:2019jts}%
  \BibitemOpen
  \bibfield  {author} {\bibinfo {author} {\bibfnamefont {R.~T.}\ \bibnamefont
  {Co}}, \bibinfo {author} {\bibfnamefont {L.~J.}\ \bibnamefont {Hall}},\ and\
  \bibinfo {author} {\bibfnamefont {K.}~\bibnamefont {Harigaya}},\ }\bibfield
  {title} {\bibinfo {title} {{Axion Kinetic Misalignment Mechanism}},\ }\href
  {https://doi.org/10.1103/PhysRevLett.124.251802} {\bibfield  {journal}
  {\bibinfo  {journal} {Phys. Rev. Lett.}\ }\textbf {\bibinfo {volume} {124}},\
  \bibinfo {pages} {251802} (\bibinfo {year} {2020}{\natexlab{a}})},\ \Eprint
  {https://arxiv.org/abs/1910.14152} {arXiv:1910.14152 [hep-ph]} \BibitemShut
  {NoStop}%
\bibitem [{\citenamefont {Chang}\ and\ \citenamefont
  {Cui}(2020)}]{Chang:2019tvx}%
  \BibitemOpen
  \bibfield  {author} {\bibinfo {author} {\bibfnamefont {C.-F.}\ \bibnamefont
  {Chang}}\ and\ \bibinfo {author} {\bibfnamefont {Y.}~\bibnamefont {Cui}},\
  }\bibfield  {title} {\bibinfo {title} {{New Perspectives on Axion
  Misalignment Mechanism}},\ }\href
  {https://doi.org/10.1103/PhysRevD.102.015003} {\bibfield  {journal} {\bibinfo
   {journal} {Phys. Rev. D}\ }\textbf {\bibinfo {volume} {102}},\ \bibinfo
  {pages} {015003} (\bibinfo {year} {2020})},\ \Eprint
  {https://arxiv.org/abs/1911.11885} {arXiv:1911.11885 [hep-ph]} \BibitemShut
  {NoStop}%
\bibitem [{\citenamefont {Co}\ \emph {et~al.}(2020{\natexlab{b}})\citenamefont
  {Co}, \citenamefont {Hall}, \citenamefont {Harigaya}, \citenamefont {Olive},\
  and\ \citenamefont {Verner}}]{Co:2020dya}%
  \BibitemOpen
  \bibfield  {author} {\bibinfo {author} {\bibfnamefont {R.~T.}\ \bibnamefont
  {Co}}, \bibinfo {author} {\bibfnamefont {L.~J.}\ \bibnamefont {Hall}},
  \bibinfo {author} {\bibfnamefont {K.}~\bibnamefont {Harigaya}}, \bibinfo
  {author} {\bibfnamefont {K.~A.}\ \bibnamefont {Olive}},\ and\ \bibinfo
  {author} {\bibfnamefont {S.}~\bibnamefont {Verner}},\ }\bibfield  {title}
  {\bibinfo {title} {{Axion Kinetic Misalignment and Parametric Resonance from
  Inflation}},\ }\href {https://doi.org/10.1088/1475-7516/2020/08/036}
  {\bibfield  {journal} {\bibinfo  {journal} {JCAP}\ }\textbf {\bibinfo
  {volume} {08}},\ \bibinfo {pages} {036}},\ \Eprint
  {https://arxiv.org/abs/2004.00629} {arXiv:2004.00629 [hep-ph]} \BibitemShut
  {NoStop}%
\bibitem [{\citenamefont {Co}\ \emph {et~al.}(2021)\citenamefont {Co},
  \citenamefont {Fernandez}, \citenamefont {Ghalsasi}, \citenamefont {Hall},\
  and\ \citenamefont {Harigaya}}]{Co:2020jtv}%
  \BibitemOpen
  \bibfield  {author} {\bibinfo {author} {\bibfnamefont {R.~T.}\ \bibnamefont
  {Co}}, \bibinfo {author} {\bibfnamefont {N.}~\bibnamefont {Fernandez}},
  \bibinfo {author} {\bibfnamefont {A.}~\bibnamefont {Ghalsasi}}, \bibinfo
  {author} {\bibfnamefont {L.~J.}\ \bibnamefont {Hall}},\ and\ \bibinfo
  {author} {\bibfnamefont {K.}~\bibnamefont {Harigaya}},\ }\bibfield  {title}
  {\bibinfo {title} {{Lepto-Axiogenesis}},\ }\href
  {https://doi.org/10.1007/JHEP03(2021)017} {\bibfield  {journal} {\bibinfo
  {journal} {JHEP}\ }\textbf {\bibinfo {volume} {03}},\ \bibinfo {pages}
  {017}},\ \Eprint {https://arxiv.org/abs/2006.05687} {arXiv:2006.05687
  [hep-ph]} \BibitemShut {NoStop}%
\bibitem [{\citenamefont {Er{\"o}ncel}\ \emph {et~al.}(2024)\citenamefont
  {Er{\"o}ncel}, \citenamefont {Sato}, \citenamefont {Servant},\ and\
  \citenamefont {S{\o}rensen}}]{Eroncel:2024rpe}%
  \BibitemOpen
  \bibfield  {author} {\bibinfo {author} {\bibfnamefont {C.}~\bibnamefont
  {Er{\"o}ncel}}, \bibinfo {author} {\bibfnamefont {R.}~\bibnamefont {Sato}},
  \bibinfo {author} {\bibfnamefont {G.}~\bibnamefont {Servant}},\ and\ \bibinfo
  {author} {\bibfnamefont {P.}~\bibnamefont {S{\o}rensen}},\ }\bibfield
  {title} {\bibinfo {title} {{Model implementations of axion dark matter from
  kinetic misalignment}},\ }\href@noop {} {\  (\bibinfo {year} {2024})},\
  \Eprint {https://arxiv.org/abs/2408.08355} {arXiv:2408.08355 [hep-ph]}
  \BibitemShut {NoStop}%
\bibitem [{\citenamefont {Nakagawa}\ \emph {et~al.}(2021)\citenamefont
  {Nakagawa}, \citenamefont {Takahashi},\ and\ \citenamefont
  {Yamada}}]{Nakagawa:2020zjr}%
  \BibitemOpen
  \bibfield  {author} {\bibinfo {author} {\bibfnamefont {S.}~\bibnamefont
  {Nakagawa}}, \bibinfo {author} {\bibfnamefont {F.}~\bibnamefont
  {Takahashi}},\ and\ \bibinfo {author} {\bibfnamefont {M.}~\bibnamefont
  {Yamada}},\ }\bibfield  {title} {\bibinfo {title} {{Trapping Effect for QCD
  Axion Dark Matter}},\ }\href {https://doi.org/10.1088/1475-7516/2021/05/062}
  {\bibfield  {journal} {\bibinfo  {journal} {JCAP}\ }\textbf {\bibinfo
  {volume} {05}},\ \bibinfo {pages} {062}},\ \Eprint
  {https://arxiv.org/abs/2012.13592} {arXiv:2012.13592 [hep-ph]} \BibitemShut
  {NoStop}%
\bibitem [{\citenamefont {Di~Luzio}\ \emph {et~al.}(2021)\citenamefont
  {Di~Luzio}, \citenamefont {Gavela}, \citenamefont {Quilez},\ and\
  \citenamefont {Ringwald}}]{DiLuzio:2021gos}%
  \BibitemOpen
  \bibfield  {author} {\bibinfo {author} {\bibfnamefont {L.}~\bibnamefont
  {Di~Luzio}}, \bibinfo {author} {\bibfnamefont {B.}~\bibnamefont {Gavela}},
  \bibinfo {author} {\bibfnamefont {P.}~\bibnamefont {Quilez}},\ and\ \bibinfo
  {author} {\bibfnamefont {A.}~\bibnamefont {Ringwald}},\ }\bibfield  {title}
  {\bibinfo {title} {{Dark matter from an even lighter QCD axion: trapped
  misalignment}},\ }\href {https://doi.org/10.1088/1475-7516/2021/10/001}
  {\bibfield  {journal} {\bibinfo  {journal} {JCAP}\ }\textbf {\bibinfo
  {volume} {10}},\ \bibinfo {pages} {001}},\ \Eprint
  {https://arxiv.org/abs/2102.01082} {arXiv:2102.01082 [hep-ph]} \BibitemShut
  {NoStop}%
\bibitem [{\citenamefont {Di~Luzio}\ and\ \citenamefont
  {S{\o}rensen}(2024)}]{DiLuzio:2024fyt}%
  \BibitemOpen
  \bibfield  {author} {\bibinfo {author} {\bibfnamefont {L.}~\bibnamefont
  {Di~Luzio}}\ and\ \bibinfo {author} {\bibfnamefont {P.}~\bibnamefont
  {S{\o}rensen}},\ }\bibfield  {title} {\bibinfo {title} {{Axion production via
  trapped misalignment from Peccei-Quinn symmetry breaking}},\ }\href
  {https://doi.org/10.1007/JHEP10(2024)239} {\bibfield  {journal} {\bibinfo
  {journal} {JHEP}\ }\textbf {\bibinfo {volume} {10}},\ \bibinfo {pages}
  {239}},\ \Eprint {https://arxiv.org/abs/2408.04623} {arXiv:2408.04623
  [hep-ph]} \BibitemShut {NoStop}%
\bibitem [{\citenamefont {Harigaya}\ and\ \citenamefont
  {Leedom}(2020)}]{Harigaya:2019qnl}%
  \BibitemOpen
  \bibfield  {author} {\bibinfo {author} {\bibfnamefont {K.}~\bibnamefont
  {Harigaya}}\ and\ \bibinfo {author} {\bibfnamefont {J.~M.}\ \bibnamefont
  {Leedom}},\ }\bibfield  {title} {\bibinfo {title} {{QCD Axion Dark Matter
  from a Late Time Phase Transition}},\ }\href
  {https://doi.org/10.1007/JHEP06(2020)034} {\bibfield  {journal} {\bibinfo
  {journal} {JHEP}\ }\textbf {\bibinfo {volume} {06}},\ \bibinfo {pages}
  {034}},\ \Eprint {https://arxiv.org/abs/1910.04163} {arXiv:1910.04163
  [hep-ph]} \BibitemShut {NoStop}%
\bibitem [{\citenamefont {Barman}\ \emph {et~al.}(2022)\citenamefont {Barman},
  \citenamefont {Bernal}, \citenamefont {Ramberg},\ and\ \citenamefont
  {Visinelli}}]{Barman:2021rdr}%
  \BibitemOpen
  \bibfield  {author} {\bibinfo {author} {\bibfnamefont {B.}~\bibnamefont
  {Barman}}, \bibinfo {author} {\bibfnamefont {N.}~\bibnamefont {Bernal}},
  \bibinfo {author} {\bibfnamefont {N.}~\bibnamefont {Ramberg}},\ and\ \bibinfo
  {author} {\bibfnamefont {L.}~\bibnamefont {Visinelli}},\ }\bibfield  {title}
  {\bibinfo {title} {{QCD Axion Kinetic Misalignment without Prejudice}},\
  }\href {https://doi.org/10.3390/universe8120634} {\bibfield  {journal}
  {\bibinfo  {journal} {Universe}\ }\textbf {\bibinfo {volume} {8}},\ \bibinfo
  {pages} {634} (\bibinfo {year} {2022})},\ \Eprint
  {https://arxiv.org/abs/2111.03677} {arXiv:2111.03677 [hep-ph]} \BibitemShut
  {NoStop}%
\bibitem [{\citenamefont {Arias}\ \emph {et~al.}(2023)\citenamefont {Arias},
  \citenamefont {Bernal}, \citenamefont {Osi{\'n}ski},\ and\ \citenamefont
  {Roszkowski}}]{Arias:2022qjt}%
  \BibitemOpen
  \bibfield  {author} {\bibinfo {author} {\bibfnamefont {P.}~\bibnamefont
  {Arias}}, \bibinfo {author} {\bibfnamefont {N.}~\bibnamefont {Bernal}},
  \bibinfo {author} {\bibfnamefont {J.~K.}\ \bibnamefont {Osi{\'n}ski}},\ and\
  \bibinfo {author} {\bibfnamefont {L.}~\bibnamefont {Roszkowski}},\ }\bibfield
   {title} {\bibinfo {title} {{Dark matter axions in the early universe with a
  period of increasing temperature}},\ }\href
  {https://doi.org/10.1088/1475-7516/2023/05/028} {\bibfield  {journal}
  {\bibinfo  {journal} {JCAP}\ }\textbf {\bibinfo {volume} {05}},\ \bibinfo
  {pages} {028}},\ \Eprint {https://arxiv.org/abs/2207.07677} {arXiv:2207.07677
  [hep-ph]} \BibitemShut {NoStop}%
\bibitem [{\citenamefont {Papageorgiou}\ \emph {et~al.}(2023)\citenamefont
  {Papageorgiou}, \citenamefont {Qu{\'\i}lez},\ and\ \citenamefont
  {Schmitz}}]{Papageorgiou:2022prc}%
  \BibitemOpen
  \bibfield  {author} {\bibinfo {author} {\bibfnamefont {A.}~\bibnamefont
  {Papageorgiou}}, \bibinfo {author} {\bibfnamefont {P.}~\bibnamefont
  {Qu{\'\i}lez}},\ and\ \bibinfo {author} {\bibfnamefont {K.}~\bibnamefont
  {Schmitz}},\ }\bibfield  {title} {\bibinfo {title} {{Axion dark matter from
  frictional misalignment}},\ }\href {https://doi.org/10.1007/JHEP01(2023)169}
  {\bibfield  {journal} {\bibinfo  {journal} {JHEP}\ }\textbf {\bibinfo
  {volume} {01}},\ \bibinfo {pages} {169}},\ \Eprint
  {https://arxiv.org/abs/2206.01129} {arXiv:2206.01129 [hep-ph]} \BibitemShut
  {NoStop}%
\bibitem [{\citenamefont {Xu}\ and\ \citenamefont {Yun}(2023)}]{Xu:2022yxr}%
  \BibitemOpen
  \bibfield  {author} {\bibinfo {author} {\bibfnamefont {L.-X.}\ \bibnamefont
  {Xu}}\ and\ \bibinfo {author} {\bibfnamefont {S.}~\bibnamefont {Yun}},\
  }\bibfield  {title} {\bibinfo {title} {{Axion free-kick misalignment
  mechanism}},\ }\href {https://doi.org/10.1103/PhysRevD.107.L091702}
  {\bibfield  {journal} {\bibinfo  {journal} {Phys. Rev. D}\ }\textbf {\bibinfo
  {volume} {107}},\ \bibinfo {pages} {L091702} (\bibinfo {year} {2023})},\
  \Eprint {https://arxiv.org/abs/2211.13074} {arXiv:2211.13074 [hep-ph]}
  \BibitemShut {NoStop}%
\bibitem [{\citenamefont {Choi}\ and\ \citenamefont
  {Schiappacasse}(2022)}]{Choi:2022btl}%
  \BibitemOpen
  \bibfield  {author} {\bibinfo {author} {\bibfnamefont {G.}~\bibnamefont
  {Choi}}\ and\ \bibinfo {author} {\bibfnamefont {E.~D.}\ \bibnamefont
  {Schiappacasse}},\ }\bibfield  {title} {\bibinfo {title} {{PBH assisted
  search for QCD axion dark matter}},\ }\href
  {https://doi.org/10.1088/1475-7516/2022/09/072} {\bibfield  {journal}
  {\bibinfo  {journal} {JCAP}\ }\textbf {\bibinfo {volume} {09}},\ \bibinfo
  {pages} {072}},\ \Eprint {https://arxiv.org/abs/2205.02255} {arXiv:2205.02255
  [hep-ph]} \BibitemShut {NoStop}%
\bibitem [{\citenamefont {Er{\"o}ncel}\ \emph {et~al.}(2022)\citenamefont
  {Er{\"o}ncel}, \citenamefont {Sato}, \citenamefont {Servant},\ and\
  \citenamefont {S{\o}rensen}}]{Eroncel:2022vjg}%
  \BibitemOpen
  \bibfield  {author} {\bibinfo {author} {\bibfnamefont {C.}~\bibnamefont
  {Er{\"o}ncel}}, \bibinfo {author} {\bibfnamefont {R.}~\bibnamefont {Sato}},
  \bibinfo {author} {\bibfnamefont {G.}~\bibnamefont {Servant}},\ and\ \bibinfo
  {author} {\bibfnamefont {P.}~\bibnamefont {S{\o}rensen}},\ }\bibfield
  {title} {\bibinfo {title} {{ALP dark matter from kinetic fragmentation:
  opening up the parameter window}},\ }\href
  {https://doi.org/10.1088/1475-7516/2022/10/053} {\bibfield  {journal}
  {\bibinfo  {journal} {JCAP}\ }\textbf {\bibinfo {volume} {10}},\ \bibinfo
  {pages} {053}},\ \Eprint {https://arxiv.org/abs/2206.14259} {arXiv:2206.14259
  [hep-ph]} \BibitemShut {NoStop}%
\bibitem [{\citenamefont {Allali}\ \emph {et~al.}(2022)\citenamefont {Allali},
  \citenamefont {Hertzberg},\ and\ \citenamefont {Lyu}}]{Allali:2022yvx}%
  \BibitemOpen
  \bibfield  {author} {\bibinfo {author} {\bibfnamefont {I.~J.}\ \bibnamefont
  {Allali}}, \bibinfo {author} {\bibfnamefont {M.~P.}\ \bibnamefont
  {Hertzberg}},\ and\ \bibinfo {author} {\bibfnamefont {Y.}~\bibnamefont
  {Lyu}},\ }\bibfield  {title} {\bibinfo {title} {{Altered axion abundance from
  a dynamical Peccei-Quinn scale}},\ }\href
  {https://doi.org/10.1103/PhysRevD.105.123517} {\bibfield  {journal} {\bibinfo
   {journal} {Phys. Rev. D}\ }\textbf {\bibinfo {volume} {105}},\ \bibinfo
  {pages} {123517} (\bibinfo {year} {2022})},\ \Eprint
  {https://arxiv.org/abs/2203.15817} {arXiv:2203.15817 [hep-ph]} \BibitemShut
  {NoStop}%
\bibitem [{\citenamefont {Cyncynates}\ and\ \citenamefont
  {Thompson}(2023)}]{Cyncynates:2023esj}%
  \BibitemOpen
  \bibfield  {author} {\bibinfo {author} {\bibfnamefont {D.}~\bibnamefont
  {Cyncynates}}\ and\ \bibinfo {author} {\bibfnamefont {J.~O.}\ \bibnamefont
  {Thompson}},\ }\bibfield  {title} {\bibinfo {title} {{Heavy QCD axion dark
  matter from avoided level crossing}},\ }\href
  {https://doi.org/10.1103/PhysRevD.108.L091703} {\bibfield  {journal}
  {\bibinfo  {journal} {Phys. Rev. D}\ }\textbf {\bibinfo {volume} {108}},\
  \bibinfo {pages} {L091703} (\bibinfo {year} {2023})},\ \Eprint
  {https://arxiv.org/abs/2306.04678} {arXiv:2306.04678 [hep-ph]} \BibitemShut
  {NoStop}%
\bibitem [{\citenamefont {Xu}(2023)}]{Xu:2023lxw}%
  \BibitemOpen
  \bibfield  {author} {\bibinfo {author} {\bibfnamefont {Y.}~\bibnamefont
  {Xu}},\ }\bibfield  {title} {\bibinfo {title} {{Constraining axion and ALP
  dark matter from misalignment during reheating}},\ }\href
  {https://doi.org/10.1103/PhysRevD.108.083536} {\bibfield  {journal} {\bibinfo
   {journal} {Phys. Rev. D}\ }\textbf {\bibinfo {volume} {108}},\ \bibinfo
  {pages} {083536} (\bibinfo {year} {2023})},\ \Eprint
  {https://arxiv.org/abs/2308.15322} {arXiv:2308.15322 [hep-ph]} \BibitemShut
  {NoStop}%
\bibitem [{\citenamefont {Lee}\ \emph {et~al.}(2024)\citenamefont {Lee},
  \citenamefont {Murai}, \citenamefont {Takahashi},\ and\ \citenamefont
  {Yin}}]{Lee:2024oaz}%
  \BibitemOpen
  \bibfield  {author} {\bibinfo {author} {\bibfnamefont {J.}~\bibnamefont
  {Lee}}, \bibinfo {author} {\bibfnamefont {K.}~\bibnamefont {Murai}}, \bibinfo
  {author} {\bibfnamefont {F.}~\bibnamefont {Takahashi}},\ and\ \bibinfo
  {author} {\bibfnamefont {W.}~\bibnamefont {Yin}},\ }\bibfield  {title}
  {\bibinfo {title} {{Bubble misalignment mechanism for axions}},\ }\href
  {https://doi.org/10.1088/1475-7516/2024/05/122} {\bibfield  {journal}
  {\bibinfo  {journal} {JCAP}\ }\textbf {\bibinfo {volume} {05}},\ \bibinfo
  {pages} {122}},\ \Eprint {https://arxiv.org/abs/2402.09501} {arXiv:2402.09501
  [hep-ph]} \BibitemShut {NoStop}%
\bibitem [{\citenamefont {Barman}\ and\ \citenamefont
  {Datta}(2024)}]{Barman:2023icn}%
  \BibitemOpen
  \bibfield  {author} {\bibinfo {author} {\bibfnamefont {B.}~\bibnamefont
  {Barman}}\ and\ \bibinfo {author} {\bibfnamefont {A.}~\bibnamefont {Datta}},\
  }\bibfield  {title} {\bibinfo {title} {{Testing axionic dark matter during
  gravitational reheating}},\ }\href
  {https://doi.org/10.1103/PhysRevD.109.095029} {\bibfield  {journal} {\bibinfo
   {journal} {Phys. Rev. D}\ }\textbf {\bibinfo {volume} {109}},\ \bibinfo
  {pages} {095029} (\bibinfo {year} {2024})},\ \Eprint
  {https://arxiv.org/abs/2312.13821} {arXiv:2312.13821 [hep-ph]} \BibitemShut
  {NoStop}%
\bibitem [{\citenamefont {Banerjee}\ and\ \citenamefont
  {Buen-Abad}(2025)}]{Banerjee:2024ykz}%
  \BibitemOpen
  \bibfield  {author} {\bibinfo {author} {\bibfnamefont {A.}~\bibnamefont
  {Banerjee}}\ and\ \bibinfo {author} {\bibfnamefont {M.~A.}\ \bibnamefont
  {Buen-Abad}},\ }\bibfield  {title} {\bibinfo {title} {{Dynamical axion
  misalignment from the Witten effect}},\ }\href
  {https://doi.org/10.1007/JHEP02(2025)078} {\bibfield  {journal} {\bibinfo
  {journal} {JHEP}\ }\textbf {\bibinfo {volume} {02}},\ \bibinfo {pages}
  {078}},\ \Eprint {https://arxiv.org/abs/2410.21369} {arXiv:2410.21369
  [hep-ph]} \BibitemShut {NoStop}%
\bibitem [{\citenamefont {Er{\"o}ncel}\ \emph {et~al.}(2025)\citenamefont
  {Er{\"o}ncel}, \citenamefont {Gouttenoire}, \citenamefont {Sato},
  \citenamefont {Servant},\ and\ \citenamefont
  {Simakachorn}}]{Eroncel:2025qlk}%
  \BibitemOpen
  \bibfield  {author} {\bibinfo {author} {\bibfnamefont {C.}~\bibnamefont
  {Er{\"o}ncel}}, \bibinfo {author} {\bibfnamefont {Y.}~\bibnamefont
  {Gouttenoire}}, \bibinfo {author} {\bibfnamefont {R.}~\bibnamefont {Sato}},
  \bibinfo {author} {\bibfnamefont {G.}~\bibnamefont {Servant}},\ and\ \bibinfo
  {author} {\bibfnamefont {P.}~\bibnamefont {Simakachorn}},\ }\bibfield
  {title} {\bibinfo {title} {{A New Source for (QCD) Axion Dark Matter
  Production: Curvature-Induced}},\ }\href@noop {} {\  (\bibinfo {year}
  {2025})},\ \Eprint {https://arxiv.org/abs/2503.04880} {arXiv:2503.04880
  [hep-ph]} \BibitemShut {NoStop}%
\bibitem [{\citenamefont {Banerjee}\ \emph {et~al.}(2025)\citenamefont
  {Banerjee}, \citenamefont {Buen-Abad},\ and\ \citenamefont
  {Hook}}]{Banerjee:2025kov}%
  \BibitemOpen
  \bibfield  {author} {\bibinfo {author} {\bibfnamefont {A.}~\bibnamefont
  {Banerjee}}, \bibinfo {author} {\bibfnamefont {M.~A.}\ \bibnamefont
  {Buen-Abad}},\ and\ \bibinfo {author} {\bibfnamefont {A.}~\bibnamefont
  {Hook}},\ }\bibfield  {title} {\bibinfo {title} {{Stacking the Deck: Gambling
  on a Light QCD Axion}},\ }\href@noop {} {\  (\bibinfo {year} {2025})},\
  \Eprint {https://arxiv.org/abs/2507.02049} {arXiv:2507.02049 [hep-ph]}
  \BibitemShut {NoStop}%
\bibitem [{\citenamefont {Lyu}\ and\ \citenamefont {Zhao}(2025)}]{Lyu:2025jge}%
  \BibitemOpen
  \bibfield  {author} {\bibinfo {author} {\bibfnamefont {K.-F.}\ \bibnamefont
  {Lyu}}\ and\ \bibinfo {author} {\bibfnamefont {Y.}~\bibnamefont {Zhao}},\
  }\bibfield  {title} {\bibinfo {title} {{QCD Axion Domain Walls from
  Super-Cooling First Order Phase Transition}},\ }\href@noop {} {\  (\bibinfo
  {year} {2025})},\ \Eprint {https://arxiv.org/abs/2506.19918}
  {arXiv:2506.19918 [hep-ph]} \BibitemShut {NoStop}%
\bibitem [{\citenamefont {Espinosa}\ \emph {et~al.}(2012)\citenamefont
  {Espinosa}, \citenamefont {Konstandin},\ and\ \citenamefont
  {Riva}}]{Espinosa:2011ax}%
  \BibitemOpen
  \bibfield  {author} {\bibinfo {author} {\bibfnamefont {J.~R.}\ \bibnamefont
  {Espinosa}}, \bibinfo {author} {\bibfnamefont {T.}~\bibnamefont
  {Konstandin}},\ and\ \bibinfo {author} {\bibfnamefont {F.}~\bibnamefont
  {Riva}},\ }\bibfield  {title} {\bibinfo {title} {{Strong Electroweak Phase
  Transitions in the Standard Model with a Singlet}},\ }\href
  {https://doi.org/10.1016/j.nuclphysb.2011.09.010} {\bibfield  {journal}
  {\bibinfo  {journal} {Nucl. Phys. B}\ }\textbf {\bibinfo {volume} {854}},\
  \bibinfo {pages} {592} (\bibinfo {year} {2012})},\ \Eprint
  {https://arxiv.org/abs/1107.5441} {arXiv:1107.5441 [hep-ph]} \BibitemShut
  {NoStop}%
\bibitem [{\citenamefont {Manna}\ and\ \citenamefont
  {Sil}(2024)}]{Manna:2023zuq}%
  \BibitemOpen
  \bibfield  {author} {\bibinfo {author} {\bibfnamefont {S.~K.}\ \bibnamefont
  {Manna}}\ and\ \bibinfo {author} {\bibfnamefont {A.}~\bibnamefont {Sil}},\
  }\bibfield  {title} {\bibinfo {title} {{Effects of electroweak symmetry
  breaking on axionlike particles as dark matter}},\ }\href
  {https://doi.org/10.1103/PhysRevD.109.095036} {\bibfield  {journal} {\bibinfo
   {journal} {Phys. Rev. D}\ }\textbf {\bibinfo {volume} {109}},\ \bibinfo
  {pages} {095036} (\bibinfo {year} {2024})},\ \Eprint
  {https://arxiv.org/abs/2311.05125} {arXiv:2311.05125 [hep-ph]} \BibitemShut
  {NoStop}%
\bibitem [{\citenamefont {Giddings}\ and\ \citenamefont
  {Strominger}(1988)}]{Giddings:1988cx}%
  \BibitemOpen
  \bibfield  {author} {\bibinfo {author} {\bibfnamefont {S.~B.}\ \bibnamefont
  {Giddings}}\ and\ \bibinfo {author} {\bibfnamefont {A.}~\bibnamefont
  {Strominger}},\ }\bibfield  {title} {\bibinfo {title} {{Loss of incoherence
  and determination of coupling constants in quantum gravity}},\ }\href
  {https://doi.org/10.1016/0550-3213(88)90109-5} {\bibfield  {journal}
  {\bibinfo  {journal} {Nucl. Phys. B}\ }\textbf {\bibinfo {volume} {307}},\
  \bibinfo {pages} {854} (\bibinfo {year} {1988})}\BibitemShut {NoStop}%
\bibitem [{\citenamefont {Coleman}(1988)}]{Coleman:1988tj}%
  \BibitemOpen
  \bibfield  {author} {\bibinfo {author} {\bibfnamefont {S.~R.}\ \bibnamefont
  {Coleman}},\ }\bibfield  {title} {\bibinfo {title} {{Why There Is Nothing
  Rather Than Something: A Theory of the Cosmological Constant}},\ }\href
  {https://doi.org/10.1016/0550-3213(88)90097-1} {\bibfield  {journal}
  {\bibinfo  {journal} {Nucl. Phys. B}\ }\textbf {\bibinfo {volume} {310}},\
  \bibinfo {pages} {643} (\bibinfo {year} {1988})}\BibitemShut {NoStop}%
\bibitem [{\citenamefont {Rey}(1989)}]{Rey:1989mg}%
  \BibitemOpen
  \bibfield  {author} {\bibinfo {author} {\bibfnamefont {S.-J.}\ \bibnamefont
  {Rey}},\ }\bibfield  {title} {\bibinfo {title} {{The Axion Dynamics in
  Wormhole Background}},\ }\href {https://doi.org/10.1103/PhysRevD.39.3185}
  {\bibfield  {journal} {\bibinfo  {journal} {Phys. Rev. D}\ }\textbf {\bibinfo
  {volume} {39}},\ \bibinfo {pages} {3185} (\bibinfo {year}
  {1989})}\BibitemShut {NoStop}%
\bibitem [{\citenamefont {Abbott}\ and\ \citenamefont
  {Wise}(1989)}]{Abbott:1989jw}%
  \BibitemOpen
  \bibfield  {author} {\bibinfo {author} {\bibfnamefont {L.~F.}\ \bibnamefont
  {Abbott}}\ and\ \bibinfo {author} {\bibfnamefont {M.~B.}\ \bibnamefont
  {Wise}},\ }\bibfield  {title} {\bibinfo {title} {{Wormholes and Global
  Symmetries}},\ }\href {https://doi.org/10.1016/0550-3213(89)90503-8}
  {\bibfield  {journal} {\bibinfo  {journal} {Nucl. Phys. B}\ }\textbf
  {\bibinfo {volume} {325}},\ \bibinfo {pages} {687} (\bibinfo {year}
  {1989})}\BibitemShut {NoStop}%
\bibitem [{\citenamefont {Akhmedov}\ \emph {et~al.}(1992)\citenamefont
  {Akhmedov}, \citenamefont {Berezhiani},\ and\ \citenamefont
  {Senjanovic}}]{Akhmedov:1992hh}%
  \BibitemOpen
  \bibfield  {author} {\bibinfo {author} {\bibfnamefont {E.~K.}\ \bibnamefont
  {Akhmedov}}, \bibinfo {author} {\bibfnamefont {Z.~G.}\ \bibnamefont
  {Berezhiani}},\ and\ \bibinfo {author} {\bibfnamefont {G.}~\bibnamefont
  {Senjanovic}},\ }\bibfield  {title} {\bibinfo {title} {{Planck scale physics
  and neutrino masses}},\ }\href {https://doi.org/10.1103/PhysRevLett.69.3013}
  {\bibfield  {journal} {\bibinfo  {journal} {Phys. Rev. Lett.}\ }\textbf
  {\bibinfo {volume} {69}},\ \bibinfo {pages} {3013} (\bibinfo {year}
  {1992})},\ \Eprint {https://arxiv.org/abs/hep-ph/9205230}
  {arXiv:hep-ph/9205230} \BibitemShut {NoStop}%
\bibitem [{\citenamefont {Kamionkowski}\ and\ \citenamefont
  {March-Russell}(1992)}]{Kamionkowski:1992mf}%
  \BibitemOpen
  \bibfield  {author} {\bibinfo {author} {\bibfnamefont {M.}~\bibnamefont
  {Kamionkowski}}\ and\ \bibinfo {author} {\bibfnamefont {J.}~\bibnamefont
  {March-Russell}},\ }\bibfield  {title} {\bibinfo {title} {{Planck scale
  physics and the Peccei-Quinn mechanism}},\ }\href
  {https://doi.org/10.1016/0370-2693(92)90492-M} {\bibfield  {journal}
  {\bibinfo  {journal} {Phys. Lett. B}\ }\textbf {\bibinfo {volume} {282}},\
  \bibinfo {pages} {137} (\bibinfo {year} {1992})},\ \Eprint
  {https://arxiv.org/abs/hep-th/9202003} {arXiv:hep-th/9202003} \BibitemShut
  {NoStop}%
\bibitem [{\citenamefont {Kallosh}\ \emph {et~al.}(1995)\citenamefont
  {Kallosh}, \citenamefont {Linde}, \citenamefont {Linde},\ and\ \citenamefont
  {Susskind}}]{Kallosh:1995hi}%
  \BibitemOpen
  \bibfield  {author} {\bibinfo {author} {\bibfnamefont {R.}~\bibnamefont
  {Kallosh}}, \bibinfo {author} {\bibfnamefont {A.~D.}\ \bibnamefont {Linde}},
  \bibinfo {author} {\bibfnamefont {D.~A.}\ \bibnamefont {Linde}},\ and\
  \bibinfo {author} {\bibfnamefont {L.}~\bibnamefont {Susskind}},\ }\bibfield
  {title} {\bibinfo {title} {{Gravity and global symmetries}},\ }\href
  {https://doi.org/10.1103/PhysRevD.52.912} {\bibfield  {journal} {\bibinfo
  {journal} {Phys. Rev. D}\ }\textbf {\bibinfo {volume} {52}},\ \bibinfo
  {pages} {912} (\bibinfo {year} {1995})},\ \Eprint
  {https://arxiv.org/abs/hep-th/9502069} {arXiv:hep-th/9502069} \BibitemShut
  {NoStop}%
\bibitem [{\citenamefont {Draper}\ \emph {et~al.}(2022)\citenamefont {Draper},
  \citenamefont {Garcia},\ and\ \citenamefont {Reece}}]{Draper:2022pvk}%
  \BibitemOpen
  \bibfield  {author} {\bibinfo {author} {\bibfnamefont {P.}~\bibnamefont
  {Draper}}, \bibinfo {author} {\bibfnamefont {I.~G.}\ \bibnamefont {Garcia}},\
  and\ \bibinfo {author} {\bibfnamefont {M.}~\bibnamefont {Reece}},\ }\bibfield
   {title} {\bibinfo {title} {{Snowmass White Paper: Implications of Quantum
  Gravity for Particle Physics}},\ }in\ \href@noop {} {\emph {\bibinfo
  {booktitle} {{Snowmass 2021}}}}\ (\bibinfo {year} {2022})\ \Eprint
  {https://arxiv.org/abs/2203.07624} {arXiv:2203.07624 [hep-ph]} \BibitemShut
  {NoStop}%
\bibitem [{\citenamefont {Cordova}\ \emph {et~al.}(2022)\citenamefont
  {Cordova}, \citenamefont {Ohmori},\ and\ \citenamefont
  {Rudelius}}]{Cordova:2022rer}%
  \BibitemOpen
  \bibfield  {author} {\bibinfo {author} {\bibfnamefont {C.}~\bibnamefont
  {Cordova}}, \bibinfo {author} {\bibfnamefont {K.}~\bibnamefont {Ohmori}},\
  and\ \bibinfo {author} {\bibfnamefont {T.}~\bibnamefont {Rudelius}},\
  }\bibfield  {title} {\bibinfo {title} {{Generalized symmetry breaking scales
  and weak gravity conjectures}},\ }\href
  {https://doi.org/10.1007/JHEP11(2022)154} {\bibfield  {journal} {\bibinfo
  {journal} {JHEP}\ }\textbf {\bibinfo {volume} {11}},\ \bibinfo {pages}
  {154}},\ \Eprint {https://arxiv.org/abs/2202.05866} {arXiv:2202.05866
  [hep-th]} \BibitemShut {NoStop}%
\bibitem [{\citenamefont {Borsanyi}\ \emph {et~al.}(2016)\citenamefont
  {Borsanyi} \emph {et~al.}}]{Borsanyi:2016ksw}%
  \BibitemOpen
  \bibfield  {author} {\bibinfo {author} {\bibfnamefont {S.}~\bibnamefont
  {Borsanyi}} \emph {et~al.},\ }\bibfield  {title} {\bibinfo {title}
  {{Calculation of the axion mass based on high-temperature lattice quantum
  chromodynamics}},\ }\href {https://doi.org/10.1038/nature20115} {\bibfield
  {journal} {\bibinfo  {journal} {Nature}\ }\textbf {\bibinfo {volume} {539}},\
  \bibinfo {pages} {69} (\bibinfo {year} {2016})},\ \Eprint
  {https://arxiv.org/abs/1606.07494} {arXiv:1606.07494 [hep-lat]} \BibitemShut
  {NoStop}%
\bibitem [{\citenamefont {Di~Luzio}\ \emph {et~al.}(2020)\citenamefont
  {Di~Luzio}, \citenamefont {Giannotti}, \citenamefont {Nardi},\ and\
  \citenamefont {Visinelli}}]{DiLuzio:2020wdo}%
  \BibitemOpen
  \bibfield  {author} {\bibinfo {author} {\bibfnamefont {L.}~\bibnamefont
  {Di~Luzio}}, \bibinfo {author} {\bibfnamefont {M.}~\bibnamefont {Giannotti}},
  \bibinfo {author} {\bibfnamefont {E.}~\bibnamefont {Nardi}},\ and\ \bibinfo
  {author} {\bibfnamefont {L.}~\bibnamefont {Visinelli}},\ }\bibfield  {title}
  {\bibinfo {title} {{The landscape of QCD axion models}},\ }\href
  {https://doi.org/10.1016/j.physrep.2020.06.002} {\bibfield  {journal}
  {\bibinfo  {journal} {Phys. Rept.}\ }\textbf {\bibinfo {volume} {870}},\
  \bibinfo {pages} {1} (\bibinfo {year} {2020})},\ \Eprint
  {https://arxiv.org/abs/2003.01100} {arXiv:2003.01100 [hep-ph]} \BibitemShut
  {NoStop}%
\bibitem [{\citenamefont {Coleman}(1977)}]{Coleman:1977py}%
  \BibitemOpen
  \bibfield  {author} {\bibinfo {author} {\bibfnamefont {S.~R.}\ \bibnamefont
  {Coleman}},\ }\bibfield  {title} {\bibinfo {title} {{The Fate of the False
  Vacuum. 1. Semiclassical Theory}},\ }\href
  {https://doi.org/10.1103/PhysRevD.16.1248} {\bibfield  {journal} {\bibinfo
  {journal} {Phys. Rev. D}\ }\textbf {\bibinfo {volume} {15}},\ \bibinfo
  {pages} {2929} (\bibinfo {year} {1977})},\ \bibinfo {note} {[Erratum:
  Phys.Rev.D 16, 1248 (1977)]}\BibitemShut {NoStop}%
\bibitem [{\citenamefont {Linde}(1981)}]{Linde:1980tt}%
  \BibitemOpen
  \bibfield  {author} {\bibinfo {author} {\bibfnamefont {A.~D.}\ \bibnamefont
  {Linde}},\ }\bibfield  {title} {\bibinfo {title} {{Fate of the False Vacuum
  at Finite Temperature: Theory and Applications}},\ }\href
  {https://doi.org/10.1016/0370-2693(81)90281-1} {\bibfield  {journal}
  {\bibinfo  {journal} {Phys. Lett. B}\ }\textbf {\bibinfo {volume} {100}},\
  \bibinfo {pages} {37} (\bibinfo {year} {1981})}\BibitemShut {NoStop}%
\bibitem [{\citenamefont {Linde}(1983)}]{Linde:1981zj}%
  \BibitemOpen
  \bibfield  {author} {\bibinfo {author} {\bibfnamefont {A.~D.}\ \bibnamefont
  {Linde}},\ }\bibfield  {title} {\bibinfo {title} {{Decay of the False Vacuum
  at Finite Temperature}},\ }\href
  {https://doi.org/10.1016/0550-3213(83)90072-X} {\bibfield  {journal}
  {\bibinfo  {journal} {Nucl. Phys. B}\ }\textbf {\bibinfo {volume} {216}},\
  \bibinfo {pages} {421} (\bibinfo {year} {1983})},\ \bibinfo {note} {[Erratum:
  Nucl.Phys.B 223, 544 (1983)]}\BibitemShut {NoStop}%
\bibitem [{\citenamefont {Guada}\ \emph {et~al.}(2020)\citenamefont {Guada},
  \citenamefont {Nemev\v{s}ek},\ and\ \citenamefont {Pintar}}]{Guada:2020xnz}%
  \BibitemOpen
  \bibfield  {author} {\bibinfo {author} {\bibfnamefont {V.}~\bibnamefont
  {Guada}}, \bibinfo {author} {\bibfnamefont {M.}~\bibnamefont
  {Nemev\v{s}ek}},\ and\ \bibinfo {author} {\bibfnamefont {M.}~\bibnamefont
  {Pintar}},\ }\bibfield  {title} {\bibinfo {title} {{FindBounce: Package for
  multi-field bounce actions}},\ }\href
  {https://doi.org/10.1016/j.cpc.2020.107480} {\bibfield  {journal} {\bibinfo
  {journal} {Comput. Phys. Commun.}\ }\textbf {\bibinfo {volume} {256}},\
  \bibinfo {pages} {107480} (\bibinfo {year} {2020})},\ \Eprint
  {https://arxiv.org/abs/2002.00881} {arXiv:2002.00881 [hep-ph]} \BibitemShut
  {NoStop}%
\bibitem [{\citenamefont {O'hare}(2020)}]{Ohare2020-gy}%
  \BibitemOpen
  \bibfield  {author} {\bibinfo {author} {\bibfnamefont {C.}~\bibnamefont
  {O'hare}},\ }\href@noop {} {\bibinfo {title} {{cajohare/AxionLimits}:
  {AxionLimits}}} (\bibinfo {year} {2020})\BibitemShut {NoStop}%
\bibitem [{\citenamefont {Grilli~di Cortona}\ \emph {et~al.}(2016)\citenamefont
  {Grilli~di Cortona}, \citenamefont {Hardy}, \citenamefont {Pardo~Vega},\ and\
  \citenamefont {Villadoro}}]{GrillidiCortona:2015jxo}%
  \BibitemOpen
  \bibfield  {author} {\bibinfo {author} {\bibfnamefont {G.}~\bibnamefont
  {Grilli~di Cortona}}, \bibinfo {author} {\bibfnamefont {E.}~\bibnamefont
  {Hardy}}, \bibinfo {author} {\bibfnamefont {J.}~\bibnamefont {Pardo~Vega}},\
  and\ \bibinfo {author} {\bibfnamefont {G.}~\bibnamefont {Villadoro}},\
  }\bibfield  {title} {\bibinfo {title} {{The QCD axion, precisely}},\ }\href
  {https://doi.org/10.1007/JHEP01(2016)034} {\bibfield  {journal} {\bibinfo
  {journal} {JHEP}\ }\textbf {\bibinfo {volume} {01}},\ \bibinfo {pages}
  {034}},\ \Eprint {https://arxiv.org/abs/1511.02867} {arXiv:1511.02867
  [hep-ph]} \BibitemShut {NoStop}%
\bibitem [{\citenamefont {Dine}\ \emph {et~al.}(1981)\citenamefont {Dine},
  \citenamefont {Fischler},\ and\ \citenamefont {Srednicki}}]{DINE1981199}%
  \BibitemOpen
  \bibfield  {author} {\bibinfo {author} {\bibfnamefont {M.}~\bibnamefont
  {Dine}}, \bibinfo {author} {\bibfnamefont {W.}~\bibnamefont {Fischler}},\
  and\ \bibinfo {author} {\bibfnamefont {M.}~\bibnamefont {Srednicki}},\
  }\bibfield  {title} {\bibinfo {title} {A simple solution to the strong cp
  problem with a harmless axion},\ }\href
  {https://doi.org/https://doi.org/10.1016/0370-2693(81)90590-6} {\bibfield
  {journal} {\bibinfo  {journal} {Physics Letters B}\ }\textbf {\bibinfo
  {volume} {104}},\ \bibinfo {pages} {199} (\bibinfo {year}
  {1981})}\BibitemShut {NoStop}%
\bibitem [{\citenamefont {Zhitnitsky}(1980)}]{Zhitnitsky:1980tq}%
  \BibitemOpen
  \bibfield  {author} {\bibinfo {author} {\bibfnamefont {A.~R.}\ \bibnamefont
  {Zhitnitsky}},\ }\bibfield  {title} {\bibinfo {title} {{On Possible
  Suppression of the Axion Hadron Interactions. (In Russian)}},\ }\href@noop {}
  {\bibfield  {journal} {\bibinfo  {journal} {Sov. J. Nucl. Phys.}\ }\textbf
  {\bibinfo {volume} {31}},\ \bibinfo {pages} {260} (\bibinfo {year}
  {1980})}\BibitemShut {NoStop}%
\bibitem [{\citenamefont {Kim}(1979)}]{Kim:1979if}%
  \BibitemOpen
  \bibfield  {author} {\bibinfo {author} {\bibfnamefont {J.~E.}\ \bibnamefont
  {Kim}},\ }\bibfield  {title} {\bibinfo {title} {{Weak Interaction Singlet and
  Strong CP Invariance}},\ }\href {https://doi.org/10.1103/PhysRevLett.43.103}
  {\bibfield  {journal} {\bibinfo  {journal} {Phys. Rev. Lett.}\ }\textbf
  {\bibinfo {volume} {43}},\ \bibinfo {pages} {103} (\bibinfo {year}
  {1979})}\BibitemShut {NoStop}%
\bibitem [{\citenamefont {Shifman}\ \emph {et~al.}(1980)\citenamefont
  {Shifman}, \citenamefont {Vainshtein},\ and\ \citenamefont
  {Zakharov}}]{Shifman:1979if}%
  \BibitemOpen
  \bibfield  {author} {\bibinfo {author} {\bibfnamefont {M.~A.}\ \bibnamefont
  {Shifman}}, \bibinfo {author} {\bibfnamefont {A.~I.}\ \bibnamefont
  {Vainshtein}},\ and\ \bibinfo {author} {\bibfnamefont {V.~I.}\ \bibnamefont
  {Zakharov}},\ }\bibfield  {title} {\bibinfo {title} {{Can Confinement Ensure
  Natural CP Invariance of Strong Interactions?}},\ }\href
  {https://doi.org/10.1016/0550-3213(80)90209-6} {\bibfield  {journal}
  {\bibinfo  {journal} {Nucl. Phys. B}\ }\textbf {\bibinfo {volume} {166}},\
  \bibinfo {pages} {493} (\bibinfo {year} {1980})}\BibitemShut {NoStop}%
\bibitem [{\citenamefont {Kahn}\ \emph {et~al.}(2016)\citenamefont {Kahn},
  \citenamefont {Safdi},\ and\ \citenamefont {Thaler}}]{Kahn:2016aff}%
  \BibitemOpen
  \bibfield  {author} {\bibinfo {author} {\bibfnamefont {Y.}~\bibnamefont
  {Kahn}}, \bibinfo {author} {\bibfnamefont {B.~R.}\ \bibnamefont {Safdi}},\
  and\ \bibinfo {author} {\bibfnamefont {J.}~\bibnamefont {Thaler}},\
  }\bibfield  {title} {\bibinfo {title} {{Broadband and Resonant Approaches to
  Axion Dark Matter Detection}},\ }\href
  {https://doi.org/10.1103/PhysRevLett.117.141801} {\bibfield  {journal}
  {\bibinfo  {journal} {Phys. Rev. Lett.}\ }\textbf {\bibinfo {volume} {117}},\
  \bibinfo {pages} {141801} (\bibinfo {year} {2016})},\ \Eprint
  {https://arxiv.org/abs/1602.01086} {arXiv:1602.01086 [hep-ph]} \BibitemShut
  {NoStop}%
\bibitem [{\citenamefont {Ouellet}\ \emph {et~al.}(2019)\citenamefont {Ouellet}
  \emph {et~al.}}]{Ouellet:2018beu}%
  \BibitemOpen
  \bibfield  {author} {\bibinfo {author} {\bibfnamefont {J.~L.}\ \bibnamefont
  {Ouellet}} \emph {et~al.},\ }\bibfield  {title} {\bibinfo {title} {{First
  Results from ABRACADABRA-10 cm: A Search for Sub-$\mu$eV Axion Dark
  Matter}},\ }\href {https://doi.org/10.1103/PhysRevLett.122.121802} {\bibfield
   {journal} {\bibinfo  {journal} {Phys. Rev. Lett.}\ }\textbf {\bibinfo
  {volume} {122}},\ \bibinfo {pages} {121802} (\bibinfo {year} {2019})},\
  \Eprint {https://arxiv.org/abs/1810.12257} {arXiv:1810.12257 [hep-ex]}
  \BibitemShut {NoStop}%
\bibitem [{\citenamefont {Alesini}\ \emph {et~al.}(2017)\citenamefont
  {Alesini}, \citenamefont {Babusci}, \citenamefont {Di~Gioacchino},
  \citenamefont {Gatti}, \citenamefont {Lamanna},\ and\ \citenamefont
  {Ligi}}]{Alesini:2017ifp}%
  \BibitemOpen
  \bibfield  {author} {\bibinfo {author} {\bibfnamefont {D.}~\bibnamefont
  {Alesini}}, \bibinfo {author} {\bibfnamefont {D.}~\bibnamefont {Babusci}},
  \bibinfo {author} {\bibfnamefont {D.}~\bibnamefont {Di~Gioacchino}}, \bibinfo
  {author} {\bibfnamefont {C.}~\bibnamefont {Gatti}}, \bibinfo {author}
  {\bibfnamefont {G.}~\bibnamefont {Lamanna}},\ and\ \bibinfo {author}
  {\bibfnamefont {C.}~\bibnamefont {Ligi}},\ }\bibfield  {title} {\bibinfo
  {title} {{The KLASH Proposal}},\ }\href@noop {} {\  (\bibinfo {year}
  {2017})},\ \Eprint {https://arxiv.org/abs/1707.06010} {arXiv:1707.06010
  [physics.ins-det]} \BibitemShut {NoStop}%
\bibitem [{\citenamefont {Alesini}\ \emph {et~al.}(2019)\citenamefont {Alesini}
  \emph {et~al.}}]{Alesini:2019nzq}%
  \BibitemOpen
  \bibfield  {author} {\bibinfo {author} {\bibfnamefont {D.}~\bibnamefont
  {Alesini}} \emph {et~al.},\ }\bibfield  {title} {\bibinfo {title} {{KLASH
  Conceptual Design Report}},\ }\href@noop {} {\  (\bibinfo {year} {2019})},\
  \Eprint {https://arxiv.org/abs/1911.02427} {arXiv:1911.02427
  [physics.ins-det]} \BibitemShut {NoStop}%
\bibitem [{\citenamefont {Alesini}\ \emph {et~al.}(2023)\citenamefont {Alesini}
  \emph {et~al.}}]{Alesini:2023qed}%
  \BibitemOpen
  \bibfield  {author} {\bibinfo {author} {\bibfnamefont {D.}~\bibnamefont
  {Alesini}} \emph {et~al.},\ }\bibfield  {title} {\bibinfo {title} {{The
  future search for low-frequency axions and new physics with the FLASH
  resonant cavity experiment at Frascati National Laboratories}},\ }\href
  {https://doi.org/10.1016/j.dark.2023.101370} {\bibfield  {journal} {\bibinfo
  {journal} {Phys. Dark Univ.}\ }\textbf {\bibinfo {volume} {42}},\ \bibinfo
  {pages} {101370} (\bibinfo {year} {2023})},\ \Eprint
  {https://arxiv.org/abs/2309.00351} {arXiv:2309.00351 [physics.ins-det]}
  \BibitemShut {NoStop}%
\bibitem [{\citenamefont {Lee}\ \emph {et~al.}(2020)\citenamefont {Lee},
  \citenamefont {Ahn}, \citenamefont {Choi}, \citenamefont {Ko},\ and\
  \citenamefont {Semertzidis}}]{Lee:2020cfj}%
  \BibitemOpen
  \bibfield  {author} {\bibinfo {author} {\bibfnamefont {S.}~\bibnamefont
  {Lee}}, \bibinfo {author} {\bibfnamefont {S.}~\bibnamefont {Ahn}}, \bibinfo
  {author} {\bibfnamefont {J.}~\bibnamefont {Choi}}, \bibinfo {author}
  {\bibfnamefont {B.~R.}\ \bibnamefont {Ko}},\ and\ \bibinfo {author}
  {\bibfnamefont {Y.~K.}\ \bibnamefont {Semertzidis}},\ }\bibfield  {title}
  {\bibinfo {title} {{Axion Dark Matter Search around 6.7 $\mu$eV}},\ }\href
  {https://doi.org/10.1103/PhysRevLett.124.101802} {\bibfield  {journal}
  {\bibinfo  {journal} {Phys. Rev. Lett.}\ }\textbf {\bibinfo {volume} {124}},\
  \bibinfo {pages} {101802} (\bibinfo {year} {2020})},\ \Eprint
  {https://arxiv.org/abs/2001.05102} {arXiv:2001.05102 [hep-ex]} \BibitemShut
  {NoStop}%
\bibitem [{\citenamefont {Semertzidis}\ \emph {et~al.}(2019)\citenamefont
  {Semertzidis} \emph {et~al.}}]{Semertzidis:2019gkj}%
  \BibitemOpen
  \bibfield  {author} {\bibinfo {author} {\bibfnamefont {Y.~K.}\ \bibnamefont
  {Semertzidis}} \emph {et~al.},\ }\bibfield  {title} {\bibinfo {title} {{Axion
  Dark Matter Research with IBS/CAPP}},\ }\href@noop {} {\  (\bibinfo {year}
  {2019})},\ \Eprint {https://arxiv.org/abs/1910.11591} {arXiv:1910.11591
  [physics.ins-det]} \BibitemShut {NoStop}%
\bibitem [{\citenamefont {Vogel}\ \emph {et~al.}(2013)\citenamefont {Vogel}
  \emph {et~al.}}]{Vogel:2013bta}%
  \BibitemOpen
  \bibfield  {author} {\bibinfo {author} {\bibfnamefont {J.~K.}\ \bibnamefont
  {Vogel}} \emph {et~al.},\ }\bibfield  {title} {\bibinfo {title} {{IAXO - The
  International Axion Observatory}},\ }in\ \href@noop {} {\emph {\bibinfo
  {booktitle} {{8th Patras Workshop on Axions, WIMPs and WISPs}}}}\ (\bibinfo
  {year} {2013})\ \Eprint {https://arxiv.org/abs/1302.3273} {arXiv:1302.3273
  [physics.ins-det]} \BibitemShut {NoStop}%
\bibitem [{\citenamefont {Armengaud}\ \emph {et~al.}(2019)\citenamefont
  {Armengaud} \emph {et~al.}}]{IAXO:2019mpb}%
  \BibitemOpen
  \bibfield  {author} {\bibinfo {author} {\bibfnamefont {E.}~\bibnamefont
  {Armengaud}} \emph {et~al.} (\bibinfo {collaboration} {IAXO}),\ }\bibfield
  {title} {\bibinfo {title} {{Physics potential of the International Axion
  Observatory (IAXO)}},\ }\href {https://doi.org/10.1088/1475-7516/2019/06/047}
  {\bibfield  {journal} {\bibinfo  {journal} {JCAP}\ }\textbf {\bibinfo
  {volume} {06}},\ \bibinfo {pages} {047}},\ \Eprint
  {https://arxiv.org/abs/1904.09155} {arXiv:1904.09155 [hep-ph]} \BibitemShut
  {NoStop}%
\bibitem [{\citenamefont {Caldwell}\ \emph {et~al.}(2017)\citenamefont
  {Caldwell}, \citenamefont {Dvali}, \citenamefont {Majorovits}, \citenamefont
  {Millar}, \citenamefont {Raffelt}, \citenamefont {Redondo}, \citenamefont
  {Reimann}, \citenamefont {Simon},\ and\ \citenamefont
  {Steffen}}]{Caldwell:2016dcw}%
  \BibitemOpen
  \bibfield  {author} {\bibinfo {author} {\bibfnamefont {A.}~\bibnamefont
  {Caldwell}}, \bibinfo {author} {\bibfnamefont {G.}~\bibnamefont {Dvali}},
  \bibinfo {author} {\bibfnamefont {B.}~\bibnamefont {Majorovits}}, \bibinfo
  {author} {\bibfnamefont {A.}~\bibnamefont {Millar}}, \bibinfo {author}
  {\bibfnamefont {G.}~\bibnamefont {Raffelt}}, \bibinfo {author} {\bibfnamefont
  {J.}~\bibnamefont {Redondo}}, \bibinfo {author} {\bibfnamefont
  {O.}~\bibnamefont {Reimann}}, \bibinfo {author} {\bibfnamefont
  {F.}~\bibnamefont {Simon}},\ and\ \bibinfo {author} {\bibfnamefont
  {F.}~\bibnamefont {Steffen}} (\bibinfo {collaboration} {MADMAX Working
  Group}),\ }\bibfield  {title} {\bibinfo {title} {{Dielectric Haloscopes: A
  New Way to Detect Axion Dark Matter}},\ }\href
  {https://doi.org/10.1103/PhysRevLett.118.091801} {\bibfield  {journal}
  {\bibinfo  {journal} {Phys. Rev. Lett.}\ }\textbf {\bibinfo {volume} {118}},\
  \bibinfo {pages} {091801} (\bibinfo {year} {2017})},\ \Eprint
  {https://arxiv.org/abs/1611.05865} {arXiv:1611.05865 [physics.ins-det]}
  \BibitemShut {NoStop}%
\bibitem [{\citenamefont {Amaro-Seoane}\ \emph {et~al.}(2017)\citenamefont
  {Amaro-Seoane} \emph {et~al.}}]{LISA:2017pwj}%
  \BibitemOpen
  \bibfield  {author} {\bibinfo {author} {\bibfnamefont {P.}~\bibnamefont
  {Amaro-Seoane}} \emph {et~al.} (\bibinfo {collaboration} {LISA}),\ }\bibfield
   {title} {\bibinfo {title} {{Laser Interferometer Space Antenna}},\
  }\href@noop {} {\  (\bibinfo {year} {2017})},\ \Eprint
  {https://arxiv.org/abs/1702.00786} {arXiv:1702.00786 [astro-ph.IM]}
  \BibitemShut {NoStop}%
\bibitem [{\citenamefont {Yagi}\ and\ \citenamefont
  {Seto}(2011)}]{Yagi:2011wg}%
  \BibitemOpen
  \bibfield  {author} {\bibinfo {author} {\bibfnamefont {K.}~\bibnamefont
  {Yagi}}\ and\ \bibinfo {author} {\bibfnamefont {N.}~\bibnamefont {Seto}},\
  }\bibfield  {title} {\bibinfo {title} {{Detector configuration of DECIGO/BBO
  and identification of cosmological neutron-star binaries}},\ }\href
  {https://doi.org/10.1103/PhysRevD.83.044011} {\bibfield  {journal} {\bibinfo
  {journal} {Phys. Rev. D}\ }\textbf {\bibinfo {volume} {83}},\ \bibinfo
  {pages} {044011} (\bibinfo {year} {2011})},\ \bibinfo {note} {[Erratum:
  Phys.Rev.D 95, 109901 (2017)]},\ \Eprint {https://arxiv.org/abs/1101.3940}
  {arXiv:1101.3940 [astro-ph.CO]} \BibitemShut {NoStop}%
\bibitem [{\citenamefont {Crowder}\ and\ \citenamefont
  {Cornish}(2005)}]{Crowder:2005nr}%
  \BibitemOpen
  \bibfield  {author} {\bibinfo {author} {\bibfnamefont {J.}~\bibnamefont
  {Crowder}}\ and\ \bibinfo {author} {\bibfnamefont {N.~J.}\ \bibnamefont
  {Cornish}},\ }\bibfield  {title} {\bibinfo {title} {{Beyond LISA: Exploring
  future gravitational wave missions}},\ }\href
  {https://doi.org/10.1103/PhysRevD.72.083005} {\bibfield  {journal} {\bibinfo
  {journal} {Phys. Rev. D}\ }\textbf {\bibinfo {volume} {72}},\ \bibinfo
  {pages} {083005} (\bibinfo {year} {2005})},\ \Eprint
  {https://arxiv.org/abs/gr-qc/0506015} {arXiv:gr-qc/0506015} \BibitemShut
  {NoStop}%
\bibitem [{\citenamefont {Corbin}\ and\ \citenamefont
  {Cornish}(2006)}]{Corbin:2005ny}%
  \BibitemOpen
  \bibfield  {author} {\bibinfo {author} {\bibfnamefont {V.}~\bibnamefont
  {Corbin}}\ and\ \bibinfo {author} {\bibfnamefont {N.~J.}\ \bibnamefont
  {Cornish}},\ }\bibfield  {title} {\bibinfo {title} {{Detecting the cosmic
  gravitational wave background with the big bang observer}},\ }\href
  {https://doi.org/10.1088/0264-9381/23/7/014} {\bibfield  {journal} {\bibinfo
  {journal} {Class. Quant. Grav.}\ }\textbf {\bibinfo {volume} {23}},\ \bibinfo
  {pages} {2435} (\bibinfo {year} {2006})},\ \Eprint
  {https://arxiv.org/abs/gr-qc/0512039} {arXiv:gr-qc/0512039} \BibitemShut
  {NoStop}%
\bibitem [{\citenamefont {Harry}\ \emph {et~al.}(2006)\citenamefont {Harry},
  \citenamefont {Fritschel}, \citenamefont {Shaddock}, \citenamefont
  {Folkner},\ and\ \citenamefont {Phinney}}]{Harry:2006fi}%
  \BibitemOpen
  \bibfield  {author} {\bibinfo {author} {\bibfnamefont {G.~M.}\ \bibnamefont
  {Harry}}, \bibinfo {author} {\bibfnamefont {P.}~\bibnamefont {Fritschel}},
  \bibinfo {author} {\bibfnamefont {D.~A.}\ \bibnamefont {Shaddock}}, \bibinfo
  {author} {\bibfnamefont {W.}~\bibnamefont {Folkner}},\ and\ \bibinfo {author}
  {\bibfnamefont {E.~S.}\ \bibnamefont {Phinney}},\ }\bibfield  {title}
  {\bibinfo {title} {{Laser interferometry for the big bang observer}},\ }\href
  {https://doi.org/10.1088/0264-9381/23/15/008} {\bibfield  {journal} {\bibinfo
   {journal} {Class. Quant. Grav.}\ }\textbf {\bibinfo {volume} {23}},\
  \bibinfo {pages} {4887} (\bibinfo {year} {2006})},\ \bibinfo {note}
  {[Erratum: Class.Quant.Grav. 23, 7361 (2006)]}\BibitemShut {NoStop}%
\bibitem [{\citenamefont {Kawamura}\ \emph {et~al.}(2006)\citenamefont
  {Kawamura} \emph {et~al.}}]{Kawamura:2006up}%
  \BibitemOpen
  \bibfield  {author} {\bibinfo {author} {\bibfnamefont {S.}~\bibnamefont
  {Kawamura}} \emph {et~al.},\ }\bibfield  {title} {\bibinfo {title} {{The
  Japanese space gravitational wave antenna DECIGO}},\ }\href
  {https://doi.org/10.1088/0264-9381/23/8/S17} {\bibfield  {journal} {\bibinfo
  {journal} {Class. Quant. Grav.}\ }\textbf {\bibinfo {volume} {23}},\ \bibinfo
  {pages} {S125} (\bibinfo {year} {2006})}\BibitemShut {NoStop}%
\bibitem [{\citenamefont {Sesana}\ \emph {et~al.}(2021)\citenamefont {Sesana}
  \emph {et~al.}}]{Sesana:2019vho}%
  \BibitemOpen
  \bibfield  {author} {\bibinfo {author} {\bibfnamefont {A.}~\bibnamefont
  {Sesana}} \emph {et~al.},\ }\bibfield  {title} {\bibinfo {title} {{Unveiling
  the gravitational universe at $\mu$-Hz frequencies}},\ }\href
  {https://doi.org/10.1007/s10686-021-09709-9} {\bibfield  {journal} {\bibinfo
  {journal} {Exper. Astron.}\ }\textbf {\bibinfo {volume} {51}},\ \bibinfo
  {pages} {1333} (\bibinfo {year} {2021})},\ \Eprint
  {https://arxiv.org/abs/1908.11391} {arXiv:1908.11391 [astro-ph.IM]}
  \BibitemShut {NoStop}%
\bibitem [{\citenamefont {Abbott}\ \emph {et~al.}(2017)\citenamefont {Abbott}
  \emph {et~al.}}]{LIGOScientific:2016wof}%
  \BibitemOpen
  \bibfield  {author} {\bibinfo {author} {\bibfnamefont {B.~P.}\ \bibnamefont
  {Abbott}} \emph {et~al.} (\bibinfo {collaboration} {LIGO Scientific}),\
  }\bibfield  {title} {\bibinfo {title} {{Exploring the Sensitivity of Next
  Generation Gravitational Wave Detectors}},\ }\href
  {https://doi.org/10.1088/1361-6382/aa51f4} {\bibfield  {journal} {\bibinfo
  {journal} {Class. Quant. Grav.}\ }\textbf {\bibinfo {volume} {34}},\ \bibinfo
  {pages} {044001} (\bibinfo {year} {2017})},\ \Eprint
  {https://arxiv.org/abs/1607.08697} {arXiv:1607.08697 [astro-ph.IM]}
  \BibitemShut {NoStop}%
\bibitem [{\citenamefont {Reitze}\ \emph {et~al.}(2019)\citenamefont {Reitze}
  \emph {et~al.}}]{Reitze:2019iox}%
  \BibitemOpen
  \bibfield  {author} {\bibinfo {author} {\bibfnamefont {D.}~\bibnamefont
  {Reitze}} \emph {et~al.},\ }\bibfield  {title} {\bibinfo {title} {{Cosmic
  Explorer: The U.S. Contribution to Gravitational-Wave Astronomy beyond
  LIGO}},\ }\href@noop {} {\bibfield  {journal} {\bibinfo  {journal} {Bull. Am.
  Astron. Soc.}\ }\textbf {\bibinfo {volume} {51}},\ \bibinfo {pages} {035}
  (\bibinfo {year} {2019})},\ \Eprint {https://arxiv.org/abs/1907.04833}
  {arXiv:1907.04833 [astro-ph.IM]} \BibitemShut {NoStop}%
\bibitem [{\citenamefont {Punturo}\ \emph {et~al.}(2010)\citenamefont {Punturo}
  \emph {et~al.}}]{Punturo:2010zz}%
  \BibitemOpen
  \bibfield  {author} {\bibinfo {author} {\bibfnamefont {M.}~\bibnamefont
  {Punturo}} \emph {et~al.},\ }\bibfield  {title} {\bibinfo {title} {{The
  Einstein Telescope: A third-generation gravitational wave observatory}},\
  }\href {https://doi.org/10.1088/0264-9381/27/19/194002} {\bibfield  {journal}
  {\bibinfo  {journal} {Class. Quant. Grav.}\ }\textbf {\bibinfo {volume}
  {27}},\ \bibinfo {pages} {194002} (\bibinfo {year} {2010})}\BibitemShut
  {NoStop}%
\bibitem [{\citenamefont {Hild}\ \emph {et~al.}(2011)\citenamefont {Hild} \emph
  {et~al.}}]{Hild:2010id}%
  \BibitemOpen
  \bibfield  {author} {\bibinfo {author} {\bibfnamefont {S.}~\bibnamefont
  {Hild}} \emph {et~al.},\ }\bibfield  {title} {\bibinfo {title} {{Sensitivity
  Studies for Third-Generation Gravitational Wave Observatories}},\ }\href
  {https://doi.org/10.1088/0264-9381/28/9/094013} {\bibfield  {journal}
  {\bibinfo  {journal} {Class. Quant. Grav.}\ }\textbf {\bibinfo {volume}
  {28}},\ \bibinfo {pages} {094013} (\bibinfo {year} {2011})},\ \Eprint
  {https://arxiv.org/abs/1012.0908} {arXiv:1012.0908 [gr-qc]} \BibitemShut
  {NoStop}%
\bibitem [{\citenamefont {Sathyaprakash}\ \emph {et~al.}(2012)\citenamefont
  {Sathyaprakash} \emph {et~al.}}]{Sathyaprakash:2012jk}%
  \BibitemOpen
  \bibfield  {author} {\bibinfo {author} {\bibfnamefont {B.}~\bibnamefont
  {Sathyaprakash}} \emph {et~al.},\ }\bibfield  {title} {\bibinfo {title}
  {{Scientific Objectives of Einstein Telescope}},\ }\href
  {https://doi.org/10.1088/0264-9381/29/12/124013} {\bibfield  {journal}
  {\bibinfo  {journal} {Class. Quant. Grav.}\ }\textbf {\bibinfo {volume}
  {29}},\ \bibinfo {pages} {124013} (\bibinfo {year} {2012})},\ \bibinfo {note}
  {[Erratum: Class.Quant.Grav. 30, 079501 (2013)]},\ \Eprint
  {https://arxiv.org/abs/1206.0331} {arXiv:1206.0331 [gr-qc]} \BibitemShut
  {NoStop}%
\bibitem [{\citenamefont {Maggiore}\ \emph {et~al.}(2020)\citenamefont
  {Maggiore} \emph {et~al.}}]{ET:2019dnz}%
  \BibitemOpen
  \bibfield  {author} {\bibinfo {author} {\bibfnamefont {M.}~\bibnamefont
  {Maggiore}} \emph {et~al.} (\bibinfo {collaboration} {ET}),\ }\bibfield
  {title} {\bibinfo {title} {{Science Case for the Einstein Telescope}},\
  }\href {https://doi.org/10.1088/1475-7516/2020/03/050} {\bibfield  {journal}
  {\bibinfo  {journal} {JCAP}\ }\textbf {\bibinfo {volume} {03}},\ \bibinfo
  {pages} {050}},\ \Eprint {https://arxiv.org/abs/1912.02622} {arXiv:1912.02622
  [astro-ph.CO]} \BibitemShut {NoStop}%
\bibitem [{\citenamefont {Garcia-Bellido}\ \emph {et~al.}(2021)\citenamefont
  {Garcia-Bellido}, \citenamefont {Murayama},\ and\ \citenamefont
  {White}}]{Garcia-Bellido:2021zgu}%
  \BibitemOpen
  \bibfield  {author} {\bibinfo {author} {\bibfnamefont {J.}~\bibnamefont
  {Garcia-Bellido}}, \bibinfo {author} {\bibfnamefont {H.}~\bibnamefont
  {Murayama}},\ and\ \bibinfo {author} {\bibfnamefont {G.}~\bibnamefont
  {White}},\ }\bibfield  {title} {\bibinfo {title} {{Exploring the early
  Universe with Gaia and Theia}},\ }\href
  {https://doi.org/10.1088/1475-7516/2021/12/023} {\bibfield  {journal}
  {\bibinfo  {journal} {JCAP}\ }\textbf {\bibinfo {volume} {12}}\bibfield
  {number} {\bibinfo  {number} { (12)},\ \bibinfo {pages} {023}},\ }\Eprint
  {https://arxiv.org/abs/2104.04778} {arXiv:2104.04778 [hep-ph]} \BibitemShut
  {NoStop}%
\bibitem [{\citenamefont {Arnold}\ and\ \citenamefont
  {Espinosa}(1993)}]{Arnold:1992rz}%
  \BibitemOpen
  \bibfield  {author} {\bibinfo {author} {\bibfnamefont {P.~B.}\ \bibnamefont
  {Arnold}}\ and\ \bibinfo {author} {\bibfnamefont {O.}~\bibnamefont
  {Espinosa}},\ }\bibfield  {title} {\bibinfo {title} {{The Effective potential
  and first order phase transitions: Beyond leading-order}},\ }\href
  {https://doi.org/10.1103/PhysRevD.47.3546} {\bibfield  {journal} {\bibinfo
  {journal} {Phys. Rev. D}\ }\textbf {\bibinfo {volume} {47}},\ \bibinfo
  {pages} {3546} (\bibinfo {year} {1993})},\ \bibinfo {note} {[Erratum:
  Phys.Rev.D 50, 6662 (1994)]},\ \Eprint {https://arxiv.org/abs/hep-ph/9212235}
  {arXiv:hep-ph/9212235} \BibitemShut {NoStop}%
\bibitem [{\citenamefont {Carrington}(1992)}]{Carrington:1991hz}%
  \BibitemOpen
  \bibfield  {author} {\bibinfo {author} {\bibfnamefont {M.~E.}\ \bibnamefont
  {Carrington}},\ }\bibfield  {title} {\bibinfo {title} {{The Effective
  potential at finite temperature in the Standard Model}},\ }\href
  {https://doi.org/10.1103/PhysRevD.45.2933} {\bibfield  {journal} {\bibinfo
  {journal} {Phys. Rev. D}\ }\textbf {\bibinfo {volume} {45}},\ \bibinfo
  {pages} {2933} (\bibinfo {year} {1992})}\BibitemShut {NoStop}%
\bibitem [{\citenamefont {Quiros}(1999)}]{Quiros:1999jp}%
  \BibitemOpen
  \bibfield  {author} {\bibinfo {author} {\bibfnamefont {M.}~\bibnamefont
  {Quiros}},\ }\bibfield  {title} {\bibinfo {title} {{Finite temperature field
  theory and phase transitions}},\ }in\ \href@noop {} {\emph {\bibinfo
  {booktitle} {{ICTP Summer School in High-Energy Physics and Cosmology}}}}\
  (\bibinfo {year} {1999})\ pp.\ \bibinfo {pages} {187--259},\ \Eprint
  {https://arxiv.org/abs/hep-ph/9901312} {arXiv:hep-ph/9901312} \BibitemShut
  {NoStop}%
\bibitem [{\citenamefont {Coleman}\ and\ \citenamefont
  {Weinberg}(1973)}]{Coleman:1973jx}%
  \BibitemOpen
  \bibfield  {author} {\bibinfo {author} {\bibfnamefont {S.~R.}\ \bibnamefont
  {Coleman}}\ and\ \bibinfo {author} {\bibfnamefont {E.~J.}\ \bibnamefont
  {Weinberg}},\ }\bibfield  {title} {\bibinfo {title} {{Radiative Corrections
  as the Origin of Spontaneous Symmetry Breaking}},\ }\href
  {https://doi.org/10.1103/PhysRevD.7.1888} {\bibfield  {journal} {\bibinfo
  {journal} {Phys. Rev. D}\ }\textbf {\bibinfo {volume} {7}},\ \bibinfo {pages}
  {1888} (\bibinfo {year} {1973})}\BibitemShut {NoStop}%
\bibitem [{\citenamefont {Caprini}\ \emph {et~al.}(2016)\citenamefont {Caprini}
  \emph {et~al.}}]{Caprini:2015zlo}%
  \BibitemOpen
  \bibfield  {author} {\bibinfo {author} {\bibfnamefont {C.}~\bibnamefont
  {Caprini}} \emph {et~al.},\ }\bibfield  {title} {\bibinfo {title} {{Science
  with the space-based interferometer eLISA. II: Gravitational waves from
  cosmological phase transitions}},\ }\href
  {https://doi.org/10.1088/1475-7516/2016/04/001} {\bibfield  {journal}
  {\bibinfo  {journal} {JCAP}\ }\textbf {\bibinfo {volume} {04}},\ \bibinfo
  {pages} {001}},\ \Eprint {https://arxiv.org/abs/1512.06239} {arXiv:1512.06239
  [astro-ph.CO]} \BibitemShut {NoStop}%
\bibitem [{\citenamefont {Athron}\ \emph {et~al.}(2024)\citenamefont {Athron},
  \citenamefont {Bal\'azs}, \citenamefont {Fowlie}, \citenamefont {Morris},\
  and\ \citenamefont {Wu}}]{Athron:2023xlk}%
  \BibitemOpen
  \bibfield  {author} {\bibinfo {author} {\bibfnamefont {P.}~\bibnamefont
  {Athron}}, \bibinfo {author} {\bibfnamefont {C.}~\bibnamefont {Bal\'azs}},
  \bibinfo {author} {\bibfnamefont {A.}~\bibnamefont {Fowlie}}, \bibinfo
  {author} {\bibfnamefont {L.}~\bibnamefont {Morris}},\ and\ \bibinfo {author}
  {\bibfnamefont {L.}~\bibnamefont {Wu}},\ }\bibfield  {title} {\bibinfo
  {title} {{Cosmological phase transitions: From perturbative particle physics
  to gravitational waves}},\ }\href
  {https://doi.org/10.1016/j.ppnp.2023.104094} {\bibfield  {journal} {\bibinfo
  {journal} {Prog. Part. Nucl. Phys.}\ }\textbf {\bibinfo {volume} {135}},\
  \bibinfo {pages} {104094} (\bibinfo {year} {2024})},\ \Eprint
  {https://arxiv.org/abs/2305.02357} {arXiv:2305.02357 [hep-ph]} \BibitemShut
  {NoStop}%
\bibitem [{\citenamefont {Guo}\ \emph {et~al.}(2021)\citenamefont {Guo},
  \citenamefont {Sinha}, \citenamefont {Vagie},\ and\ \citenamefont
  {White}}]{Guo:2020grp}%
  \BibitemOpen
  \bibfield  {author} {\bibinfo {author} {\bibfnamefont {H.-K.}\ \bibnamefont
  {Guo}}, \bibinfo {author} {\bibfnamefont {K.}~\bibnamefont {Sinha}}, \bibinfo
  {author} {\bibfnamefont {D.}~\bibnamefont {Vagie}},\ and\ \bibinfo {author}
  {\bibfnamefont {G.}~\bibnamefont {White}},\ }\bibfield  {title} {\bibinfo
  {title} {{Phase Transitions in an Expanding Universe: Stochastic
  Gravitational Waves in Standard and Non-Standard Histories}},\ }\href
  {https://doi.org/10.1088/1475-7516/2021/01/001} {\bibfield  {journal}
  {\bibinfo  {journal} {JCAP}\ }\textbf {\bibinfo {volume} {01}},\ \bibinfo
  {pages} {001}},\ \Eprint {https://arxiv.org/abs/2007.08537} {arXiv:2007.08537
  [hep-ph]} \BibitemShut {NoStop}%
\bibitem [{\citenamefont {Caprini}\ \emph {et~al.}(2020)\citenamefont {Caprini}
  \emph {et~al.}}]{Caprini:2019egz}%
  \BibitemOpen
  \bibfield  {author} {\bibinfo {author} {\bibfnamefont {C.}~\bibnamefont
  {Caprini}} \emph {et~al.},\ }\bibfield  {title} {\bibinfo {title} {{Detecting
  gravitational waves from cosmological phase transitions with LISA: an
  update}},\ }\href {https://doi.org/10.1088/1475-7516/2020/03/024} {\bibfield
  {journal} {\bibinfo  {journal} {JCAP}\ }\textbf {\bibinfo {volume} {03}},\
  \bibinfo {pages} {024}},\ \Eprint {https://arxiv.org/abs/1910.13125}
  {arXiv:1910.13125 [astro-ph.CO]} \BibitemShut {NoStop}%
\bibitem [{\citenamefont {Hindmarsh}\ \emph {et~al.}(2021)\citenamefont
  {Hindmarsh}, \citenamefont {L\"uben}, \citenamefont {Lumma},\ and\
  \citenamefont {Pauly}}]{Hindmarsh:2020hop}%
  \BibitemOpen
  \bibfield  {author} {\bibinfo {author} {\bibfnamefont {M.~B.}\ \bibnamefont
  {Hindmarsh}}, \bibinfo {author} {\bibfnamefont {M.}~\bibnamefont {L\"uben}},
  \bibinfo {author} {\bibfnamefont {J.}~\bibnamefont {Lumma}},\ and\ \bibinfo
  {author} {\bibfnamefont {M.}~\bibnamefont {Pauly}},\ }\bibfield  {title}
  {\bibinfo {title} {{Phase transitions in the early universe}},\ }\href
  {https://doi.org/10.21468/SciPostPhysLectNotes.24} {\bibfield  {journal}
  {\bibinfo  {journal} {SciPost Phys. Lect. Notes}\ }\textbf {\bibinfo {volume}
  {24}},\ \bibinfo {pages} {1} (\bibinfo {year} {2021})},\ \Eprint
  {https://arxiv.org/abs/2008.09136} {arXiv:2008.09136 [astro-ph.CO]}
  \BibitemShut {NoStop}%
\bibitem [{\citenamefont {Grojean}\ and\ \citenamefont
  {Servant}(2007)}]{Grojean:2006bp}%
  \BibitemOpen
  \bibfield  {author} {\bibinfo {author} {\bibfnamefont {C.}~\bibnamefont
  {Grojean}}\ and\ \bibinfo {author} {\bibfnamefont {G.}~\bibnamefont
  {Servant}},\ }\bibfield  {title} {\bibinfo {title} {{Gravitational Waves from
  Phase Transitions at the Electroweak Scale and Beyond}},\ }\href
  {https://doi.org/10.1103/PhysRevD.75.043507} {\bibfield  {journal} {\bibinfo
  {journal} {Phys. Rev. D}\ }\textbf {\bibinfo {volume} {75}},\ \bibinfo
  {pages} {043507} (\bibinfo {year} {2007})},\ \Eprint
  {https://arxiv.org/abs/hep-ph/0607107} {arXiv:hep-ph/0607107} \BibitemShut
  {NoStop}%
\bibitem [{\citenamefont {Vaskonen}(2017)}]{Vaskonen:2016yiu}%
  \BibitemOpen
  \bibfield  {author} {\bibinfo {author} {\bibfnamefont {V.}~\bibnamefont
  {Vaskonen}},\ }\bibfield  {title} {\bibinfo {title} {{Electroweak
  baryogenesis and gravitational waves from a real scalar singlet}},\ }\href
  {https://doi.org/10.1103/PhysRevD.95.123515} {\bibfield  {journal} {\bibinfo
  {journal} {Phys. Rev. D}\ }\textbf {\bibinfo {volume} {95}},\ \bibinfo
  {pages} {123515} (\bibinfo {year} {2017})},\ \Eprint
  {https://arxiv.org/abs/1611.02073} {arXiv:1611.02073 [hep-ph]} \BibitemShut
  {NoStop}%
\bibitem [{\citenamefont {Espinosa}\ \emph {et~al.}(2010)\citenamefont
  {Espinosa}, \citenamefont {Konstandin}, \citenamefont {No},\ and\
  \citenamefont {Servant}}]{Espinosa:2010hh}%
  \BibitemOpen
  \bibfield  {author} {\bibinfo {author} {\bibfnamefont {J.~R.}\ \bibnamefont
  {Espinosa}}, \bibinfo {author} {\bibfnamefont {T.}~\bibnamefont
  {Konstandin}}, \bibinfo {author} {\bibfnamefont {J.~M.}\ \bibnamefont {No}},\
  and\ \bibinfo {author} {\bibfnamefont {G.}~\bibnamefont {Servant}},\
  }\bibfield  {title} {\bibinfo {title} {{Energy Budget of Cosmological
  First-order Phase Transitions}},\ }\href
  {https://doi.org/10.1088/1475-7516/2010/06/028} {\bibfield  {journal}
  {\bibinfo  {journal} {JCAP}\ }\textbf {\bibinfo {volume} {06}},\ \bibinfo
  {pages} {028}},\ \Eprint {https://arxiv.org/abs/1004.4187} {arXiv:1004.4187
  [hep-ph]} \BibitemShut {NoStop}%
\bibitem [{\citenamefont {Hindmarsh}\ \emph {et~al.}(2015)\citenamefont
  {Hindmarsh}, \citenamefont {Huber}, \citenamefont {Rummukainen},\ and\
  \citenamefont {Weir}}]{Hindmarsh:2015qta}%
  \BibitemOpen
  \bibfield  {author} {\bibinfo {author} {\bibfnamefont {M.}~\bibnamefont
  {Hindmarsh}}, \bibinfo {author} {\bibfnamefont {S.~J.}\ \bibnamefont
  {Huber}}, \bibinfo {author} {\bibfnamefont {K.}~\bibnamefont {Rummukainen}},\
  and\ \bibinfo {author} {\bibfnamefont {D.~J.}\ \bibnamefont {Weir}},\
  }\bibfield  {title} {\bibinfo {title} {{Numerical simulations of acoustically
  generated gravitational waves at a first order phase transition}},\ }\href
  {https://doi.org/10.1103/PhysRevD.92.123009} {\bibfield  {journal} {\bibinfo
  {journal} {Phys. Rev. D}\ }\textbf {\bibinfo {volume} {92}},\ \bibinfo
  {pages} {123009} (\bibinfo {year} {2015})},\ \Eprint
  {https://arxiv.org/abs/1504.03291} {arXiv:1504.03291 [astro-ph.CO]}
  \BibitemShut {NoStop}%
\end{thebibliography}%
\end{document}